\documentstyle[12pt,aasms4,epsfig]{article}
\input psfig.sty
\input psfig
\def\baselinestretch{1.0}  
\received{2001 April 9}
\accepted{2001 July 11}
\lefthead{Yuan et al.}
\righthead{Multicolor photometry of Abell 2634}

\begin{document}
~~~

\vskip 3cm

\begin{center}
\title{\LARGE\bf  Multicolor photometry of the galaxies \\
   in the central region of Abell 2634 }  
\end{center}

\vskip 2cm
\author{\large Qirong Yuan\altaffilmark{1,2,3},
 Xu Zhou\altaffilmark{1}, Jiansheng Chen\altaffilmark{1}, 
 Zhaoji Jiang\altaffilmark{1}, \\
 Jun Ma\altaffilmark{1}, Hong Wu\altaffilmark{1}, 
 Suijian Xue\altaffilmark{1}, and Jin Zhu\altaffilmark{1}}



\altaffiltext{1}{National Astronomical Observatories, Chinese Academy of
Sciences, Beijing 100012, China; zhouxu,chen@vega.bac.pku.edu.cn}

\altaffiltext{2}{Department of Physics, Nanjing Normal University,
NingHai Road 122, Nanjing 210097, China; qirongwy@public1.ptt.js.cn}

\altaffiltext{3}{Visiting astronomer, National Optical Astronomy 
Observatories, Tucson, AZ 85726, USA, under contract to China Scholarship 
Council; yuan@noao.edu} 

\vskip 1cm

\begin{abstract}

An optical photometric observation with the Beijing-Arizona-Taiwan-Connecticut 
(BATC) multicolor system 
is carried out for the central region of the nearby cluster of galaxies, Abell
2634. From the $2k \times 2k$ CCD images with fourteen filters which covers a
range of wavelength from 3600 \AA ~to 10000 \AA,  5572 sources are detected
down to $V \sim 20$ mag in a field of 56' $\times$ 56' centered on this
regular cluster of galaxies. As a result, we achieved the spectral energy
distributions (SEDs) of all sources detected. There are 178 previously known
galaxies included in our observations, 147 of which have known radial
velocities in literature. After excluding the foreground and background
galaxies, a sample of 124 known members is formed for an investigation of
the SED properties. The comparison of observed SEDs of the early-type member
galaxies with the template SEDs demonstrates the accuracy and reliability of
our photometric measurements. 

Based on the knowledge of SED properties of member galaxies, we performed the
selection of faint galaxies belonging to Abell 2634. It is well shown that the
color-color diagrams are powerful in the star/galaxy separation. As a result,
359 faint galaxies are selected by their color features.  The technique of
photometric redshift and color-magnitude correlation for the early-type
galaxies are applied for these faint galaxies, and a list of 74 faint 
member galaxies is achieved. On the basis of the  newly-generated  sample of
member galaxies, the spatial distribution and color-magnitude relation of the
galaxies in core region of Abell 2634 are discussed. There exists a
tendency that the color index dispersion of the early-type members is
larger for the outer region, which might reflect some clues about the 
environmental effect on the evolution of galaxies in a cluster. 

\end{abstract}

\keywords{galaxies: clusters: individual (Abell 2634) --- galaxies:
distances and redshifts --- galaxies: fundamental parameters
} 

\section{Introduction}


It is widely appreciated that the observations of galaxy clusters can provide
useful constraints on the theories of formation and evolution of large-scale
structure, since they are the largest gravitationally bound systems in the
Universe. A great volume of observational data has been achieved for the
galaxies belonging to a cluster, particularly for the nearby clusters of
galaxies. The rich cluster of galaxies, Abell 2634, is one example. 
Extensive observational efforts are being made to expand the kinematic
database for Abell 2634 because it  has many intriguing observational features.
First, its central galaxy,  NGC 7720, classified as cD first by Matthews,
Morgan, \& Schmidt (1964) and then classified as D galaxy by Dressler (1980a),
was found to have a  companion galaxy, namely NGC 7720A, with a projected
distance of 7.2  $h_{75}^{-1}$ kpc and a velocity difference of about 1000 km
s$^{-1}$.  Second, a surprisingly large negative peculiar velocity for
Abell 2634 was derived when the fundamental plane (FP) relation   $D_n -
\sigma$ (Dressler et al. 1987) was sketchily extended to estimate the
distance of Abell 2634  (Lucey et al. 1991), which is challengeable for the
universality of the FP and Tully-Fisher (TF) relations. With the effort for
several years by Lucey et al. (1997) and Scodeggio, Giovanelli \& Haynes
(1997), larger samples of member galaxies with more accurate photometric
measurements (including $I$ band observations) are constructed. Abell 2634 was
found to have a negligibly small peculiar velocity with respect to the cosmic
microwave background reference frame. The FP  distance estimate of Abell
2634 is in good agreement with that given by the TF relation. 

Additionally, the prototypical wide-angle tailed (WAT) radio source, 3C465, 
is found to coincide with the central cD galaxy, NGC 7720 (Griffen 1963; 
Eilek et al. 1984; O'Donoghue, Owen \& Eilek 1990). A study of this cluster 
by Pinkney et al.(1993) suggested that the large scale
turbulent gas motions fueled by a cluster-subcluster merging process might be
responsible for the observed bending of radio tails.  To investigate issues
related to the kinematics of NGC 7720, Scodeggio et al. (1995; hereafter
SSGH) carried out the multi-fiber spectroscopy and 21 cm line
observations, and particularly analyzed the structure, kinematics, and
morphological segregation of galaxies within the central half degree region.
There are 99 member galaxies bright than $m_v \sim 16.0$ in the inner
$0.^{\circ}5$ region, which is called the HDR (half degree region) sample.  
The completeness of
the inner region of one square degree reached $m_{pg} \sim 16.0$. As a
result, the morphological dependence of spatial and kinematic properties was
found: the early-type galaxies appear to be a relaxed system, while the
spirals have much larger velocity dispersion.

Studies of the luminosity function of cluster galaxies show that most 
intrinsically bright galaxies in clusters, $M_B < -16$, are giant ellipticals.
Down to $m_v \sim 18.5$ in Abell 2634, the majority of apparently faint
member galaxies are still intrinsically bright. There should be many giant
ellipticals within the apparent magnitude range of $16.0 < m_v < 18.5$. In
order to enrich our understanding of the structure and dynamics of Abell 2634,
these faint member galaxies should be taken into account. However, the
spectroscopic observations are not available yet for a great number of
galaxies fainter than 16.0 mag in Abell 2634.
The multicolor optical photometric observation therefore becomes a
common alternative means to study some properties of the member galaxies, such
as the color-magnitude relation (Bower, Lucey \& Ellis 1992),
Butcher-Oemler effect (Butcher \& Oemler 1978) and morphology-density relation
(Dressler 1980b).

Due to a large field of view, the Schmidt Telescope (1-meter class) telescope
equipped with a large format CCD, is one of most suitable facilities for 
obtaining photometric data for the nearby clusters of galaxies. The
multicolor photometry can provide the spectral energy distributions (SEDs) of
all the objects within the field of view. Comparing the integration times
required for spectroscopy to a certain depth on a large telescope, the amount
of observing times for multicolor photometry are much shorter and therefore
more readily scheduled. The Beijing-Arizona-Connecticut (BATC) multicolor
photometric survey is designed to obtain the SED information of cluster 
galaxies and other types of objects (Fan et al. 1996). This paper will
present the multicolor optical  photometry of the central region of Abell
2634, using the 60/90 cm Schmidt Telescope of Beijing Astronomical
Observatory (BAO) with 14 BATC filters covering a wide wavelength range
from 3800 \AA ~to nearly 10000 \AA.  The comparison of observed SEDs for
the previously known member galaxies with the template SEDs demonstrates the
accuracy and reliability of our photometric measurements. Our multicolor
photometry contains a large quantity of bright member galaxies with
sufficiently small photometric errors, which allows a verification of
color-magnitude correlation as well as a testing application of the
technique of photometric redshift. After performing a reliable star/galaxy
separation by the color-color diagrams, we carried out a membership
selection based on the SED features of known member galaxies. The technique
of photometric redshift and color-magnitude correlation have provided useful
constraints on the membership selection. As a result, we cataloged the SEDs
of the newly-selected faint members, based on which some properties of the
enlarged sample of member galaxies can be addressed. We believe that current
study will not only supplement the data bases of member galaxies in this
well-studied cluster, but also extend the investigations on the spatial
distribution and color-magnitude relation of the galaxies in the core of
Abell 2634 to an unprecedented depth. 

This paper is structured as follows. 
In Section 2, we will present our observations and data reduction, as well 
as the photometric catalogs of the objects (including the known member
galaxies) in the central field of Abell 2634.  The analysis of the SEDs for the
known member galaxies is shown in Section 3.  A SED selection of faint member
galaxies is given in Section 4, and the spatial distribution and
color properties of the enlarged sample of member galaxies is
presented in Section 5. Finally, we will give a summary in Section 6.  The
cosmological parameters $H_0=50$ km s$^{-1}$ Mpc$^{-1}$ and $q_0=0.5$ have
been assumed.

\section {Observations and data reduction}
\subsection{BATC observations}

Abell 2634 is a nearby ($z \sim 0.03$) regular cluster of galaxies,
classified by Abell (1958) as of richness class 1. Dressler (1980a) listed 
132 galaxies in his catalog, and the fraction of early-type galaxies (E/S0) 
is about 63\%. In the HDR sample of SSGH, the fraction of early-type galaxies 
reaches $\sim$ 71\%. This cluster appears to be at a distance of
$\sim$ 9000 km s$^{-1}$, projected on the main ridge of the Pisces-Perseus
Supercluster (PPS) (see Fig.1 in Wegner, Haynes, \& Giovanelli 1993).
Its companion Abell 2666 is closely located at approximately 3 degree to the
east, with a heliocentric redshift of $\sim$ 8154 km s$^{-1}$ 
(Struble \& Rood 1999). 

Our multicolor optical photometry concentrates on a region of 56 $\times$
56 arcmin$^2$ centered on NGC 7720. The 69/90 cm f/3 Schmidt Telescope of
BAO, located at Xinglong site with
an altitude of 900m, was used to obtain the photometric observations. A Ford
2048 $\times$ 2048 CCD camera was equipped at the prime focus of the
telescope. The field of view was 0.95 square degree and the spatial scale was
$1.67''$ per pixel. The details of the Schmidt Telescope, CCD camera and
data-acquisition system can be found elsewhere (Fan et al. 1996; Yan et al.
2000). We make use of the BATC filter system which includes 16
intermediate-band filters covering the whole optical wavelength range from
$\sim$ 3000 \AA ~to 10000 \AA. The transmissions of these filters can been 
seen in Fig. 1 of Zhou et al. (2001).

Only 14 filters, from $b$ to $p$, were used for the observations of Abell
2634. The filter number, filter name, effective wavelength centroid and FWHM
for each filter are listed in Table 1. In total, we have made more than 40
hours exposures, and obtained 214 images on the central region of Abell
2634. After a check of the image quality, 173 images with nearly 36
hours of exposure were selected to be combined.  The combined images cover an
area defined by a right ascension range from $23^h 36^m 16^s$ to $23^h 40^m
23^s$, and a declination range from $26^{\circ} 33' 30''$ to $27^{\circ} 30'
01''$ (for equinox of 2000.0).  The seeing of the combined images were quite
different from band to band, but the typical seeing of combined image for
each filter band was about 5 arcsec.   

Almost all nights were thought to be photometric by the observers. The
standard  stars were observed  between  air-masses 1  to 2 for each
program filter band. Normally, four to six filters were selected to make
flux calibrations during each photometric night. The standard stars were put
near the CCD center and only 300 $\times$ 300 sub-images were taken for the
standard stars to save readout time and disk space. The extinction 
coefficient and magnitude zero point obtained from standard stars were used
for making the flux calibration on the BATC field images. The details of
the observation for calibration are described in Yan et al. (2000). Table 1
also gives the parameters for our observations.

\subsection{Data reduction}

The bias subtractions and dome flat-field corrections were done on the CCD 
images. The cosmic ray and bad pixel effect were corrected by comparing
the images. The images were re-centered  and  position calibration was 
performed by using the Guide Star Catalog (GSC). The fluxes of 
intermediate-band filter images  were calibrated by Oke-Gunn standard stars
(Oke \& Gunn 1983; Fukugita et al. 1996). 

As in other works on this image field, we are preparing the spectral energy  
distribution (SED) catalog of sources. The magnitudes
were measured by the aperture photometry with detection threshold of
$4\sigma$ of the sky fluctuation per pixel. A minimum of 1 pixel above the
threshold was required for detection. This might introduce a false detection,
particularly for the nearby very bright stars. 
Since most of galaxies in the images are obviously  extended,
we finally selected a fixed
aperture of 5 pixels to do the photometry, which is large enough to
make different seeing effects negligible. Although the resulting magnitudes
given by photometry with fixed aperture are not the same as the total
magnitudes of galaxies in literatures, this is a proper way to obtain the 
reliable color indices (i.e., relative SEDs) of all the sources. The
photometry  program developed  by Bertin \& Arnouts (1996), SExtractor
(Ver2.1.6) namely, was used for the photometry. During extraction of the
sources, a filter of $3\times3$ ``all-ground'' convolution mask with FWHM =
2 pixels was used. Since the adopted aperture size is rather large,
contamination of the light by nearby objects may sometimes not be 
negligible. The number of sources detected in the combined images are shown
in Table 1.

We used four standard stars, namely, BD+17d4708, BD+26d2606, HD19445, and
HD84937, in Oke \& Gunn (1983) for the BATC flux calibration. The absolute 
fluxes of these stars were taken from Fukugita et al. (1996). The definitions
of BATC magnitude are in the AB$_{\nu}$ system of Oke \& Gunn (1983):$ m_{\rm
BATC} = -2.5\cdot {\rm log} \widetilde{F_{\nu}} - 48.6. $
Once the aperture photometry on the images of standard stars was done,
the extinction coefficients, $K$, and magnitude zero points, $C$, can be
obtained by fitting the extinction curve of magnitude {\it versus} airmass:
$
 m_{inst}-m_{BATC} = K\cdot X + C,
$
where $X$ is the airmass of the image.  $K$ and $C$ were derived by a median
fitting of the data points with straight line. The procedures of flux
calibration  are detailed in Zhou et al. (2001). 

For $c$ and $o$ filters, we do not have the calibration images. Fortunately,
we developed a method to calibrate the SEDs of objects for our large field
multicolor photometric system, based on the the SED library (Zhou et al.
1999), which is the so-called model calibration. This method heavily improved
the quality of the observations taken under not so perfect photometric
condition. The SED calibration was done for our combined images. Then, we
can derive the calibrated magnitudes in $c$ and $o$ bands from the
information of the calibrated SEDs. For other filter bands, the magnitudes
derived by ordinary standard calibrations were adopted for further analyses.

\subsection{Photometric catalog}

We achieved the photometric magnitudes within 14 filter bands for 5572 sources
in the central region (56' $\times$ 56') of Abell 2634. This SED catalog
will be electronically provided upon request. To cross-identify
all the known galaxies within the our field, we made use of the NASA/IPAC
Extragalactic Database (NED). As a result, 178 known galaxies were found
to have counterparts in our SED catalog. The identification is unambiguous 
for a majority of these galaxies, according to their positional offsets and
apparent magnitudes. The SEDs of these 178 galaxies are cataloged in Table 2,
which structured as follows:

\begin{tabular}{ll}  
Column 1: & Number of the galaxies, sorted by the R.A.in 2000 epoch.\\
Column 2,3: & R.A. and declination in 2000 epoch, in ``hhmmss.ss'' and
``ddmmss.ss'' \\
&mode, given by our photometric measurements. \\
Column 4: & Redshift, in km s$^{-1}$, provided by the NED.\\
Column 5: & Membership code: member galaxies are identified as ``m'';
foreground \\
& galaxies are marked as ``f''; background galaxies are marked by ``b''; \\ 
& and ``?'' means possible members. See text for details. \\
Column 6-19: & Photometric magnitudes in $b - p$ filter bands. The value of 0.0
means\\ 
& no detection in this filter band. \\
\end{tabular}

It should be noted that the photometric magnitudes given in Table 2 might be
somewhat different from the total magnitudes of galaxies given in some
catalogs, since we used a fixed aperture to do the photometry. What we
attempted to obtain was the relative SEDs of objects in the central region
of  Abell 2634. 


We checked two kinds of errors respectively given by the photometric program
SExtractor and by a comparison with the stellar spectral templates of Gunn \&
Stryker (1983). The statistic errors given by the SExtractor are smaller than
the real measurement errors. We compared the errors using different images
with the same filter. By separating stars into different sub-groups of
magnitudes with a interval of 0.5 mag, the mean measurement errors at
specified magnitudes were derived. We found that the measurement errors for
each filter band are larger at fainter depth. Typically, the errors are
$0.02^m$ for bright stars (say, $m<16.5^m$), and $0.05^m$ at $m=19.0^m$. 

For magnitudes in $c$ and $o$ filters, the errors from the model calibration
(Zhou et al. 1999) should be taken into account. By comparing the
observed SEDs and the template SEDs, the typical errors are $0.05^m$ and 
$0.10^m$ respectively for $c$ and $o$ filters, which should be larger than
the upper limits of the observational errors because the difference between
observed SEDs and template SEDs includes observational measurements errors,
template SED errors, and data fitting errors. The principle of this kind
of error estimation can be found in Zhou et al. (2001).

\section{ Analyses of the SEDs of the known member galaxies }
\subsection{Sample of 124 known member galaxies}


In general, it is not reasonable to select all the member galaxies in a 
nearby cluster by simply setting two fixed velocity limits because the 
dispersion of line-of-sight velocities of members is expected to be larger in
the inner region of cluster and smaller in the outer region, whether the
cluster is gravitationally bound and isolated or whether there exists a
significant amount of secondary infall. The velocity dispersion as a function
of angular separation for all galaxies within central 6$^\circ$ region of
Abell 2634 was calculated by SSGH (see Fig. 5 in SSGH). 

Since our observations focus on only 0.95 square degree centered on Abell 2634, 
it is reasonable to select member galaxies within our field by the 
criterion of 7000  km s$^{-1} < cz <$ 13000 km s$^{-1}$. The velocity
dispersion in central region of Abell 2634 should be larger, and a negligible
contamination of galaxies in Abell 2666 (with $3^\circ$ angular separation)
can be expected.  Among 178 known galaxies, there are 147 galaxies having
redshift information, including 124 member galaxies (i.e., 7000 km s$^{-1} < cz
<$ 13000 km s$^{-1}$), 22 background galaxies (i.e., $cz > 13000$ km
s$^{-1}$), and 1 foreground galaxy (i.e., $cz < 7000$ km s$^{-1}$). 
A check of our list of member galaxies with those in SSGH shows that all 
galaxies in the HDR sample are found to be included.  Fig.1 shows the 
distribution of all galaxies with known redshifts in our field.
Some galaxies not belonging to Abell 2634 are found in our field,
including 7 galaxies belonging to a rich cluster Abell 2622, centered at
$0.^{\circ}9$ to the northeast of Abell 2634, and 10 galaxies belonging to the
a background cluster which is almost behind the center of Abell 2634. One
galaxy with redshift 5609 km s$^{-1}$ should belong to the foreground branch
of the PPS (Pinkney et al. 1993).  It is easy to see that the highest peak 
occurs at $\sim 9300$ km s$^{-1}$, which is associated with Abell 2634.

\vskip 5mm
\centerline{\framebox[10cm][c]{\Large \bf Here is Figure 1 (a)(b) }}


\subsection{Comparison between SED templates and observed SEDs}

In order to check the reliability of our SEDs, it is interesting and
important to compare the SEDs of 124 member galaxies with the template SEDs.
To obtain the template SEDs for all types of galaxies, we shifted the 
original galaxy library spectra (Kinney et al. 1996) with a redshift of $z =
0.03$, the mean redshift of Abell 2634, and then convolved them with the
filter transmission curves of BATC filter system. The majority of galaxies in
the central region are early-type galaxies (i.e., E and S0). To form
a sample of early-type member galaxies, we searched the morphological
information in the Principle Catalog of Galaxies (PGC) (version 1999) in 
the Lyon-Meudon Extragalactic Database (LEDA) (Paturel et al. 1997). 
Then, we checked and supplemented the list, referring to some recent
literature (Scodeggio et al. 1998; Lucey et al. 1997; Scodeggio et al.
1997; Pinkey et al. 1993; Lucey et al. 1991). As a result, 60 early-type
galaxies (E and S0), 2 S0a  galaxies, 9 spirals and 1 irregular galaxy are
found in our sample. Note that it is difficult to separate lenticulars from
ellipticals from the previous imaging and spectroscopic observations. That
is why the morphological indices given in the literature for certain number
of early-type galaxies are quite diverse and inconsistent. Furthermore, the
template SEDs of our BATC filter system for ellipticals and lenticulars are
found to be almost identical. We shall analyze the SED features of
early-type galaxies, including E and S0, as a whole in current study.

The SEDs of 60 early-type galaxies with reference to the magnitude in the $h$
band  are shown in Fig.2a.  We find that their SEDs are quite similar. The
negligible SED dispersion is rather reasonable, which should be mainly due to
the slight deviation of their redshifts and stellar metallicities. The mean
SED of our sample has a better fit to the SED template, which can be seen in
Fig.2(b).

\vskip 5mm

\centerline{\framebox[12cm][c]{\Large \bf Here is Figure 2 (a) (b) }}
\vskip 5mm

We find that many observed SEDs of galaxies are very similar to the template
spectra. As an example, Fig. 3 show the SED of the central galaxy, NGC 7720,
and the template spectrum of ellipticals.

\centerline{\framebox[10cm][c]{\Large \bf Here is Figure 3}}


\subsection{A verification of color-magnitude relation}

Our multicolor photometric observations allow a verification of the so-called
color-magnitude relation. Many previous studies showed that fainter
early-type galaxies tend to have colors bluer than the brighter galaxies do
(see Bower, Lucey, \& Ellis 1992 and references therein). In other word, there
should be a correlation for the early-type galaxies between their colors and
absolute magnitudes. For the members of a galaxy cluster, the differences in
apparent magnitudes can well reflect the differences in absolute luminosities.

Figure 4 shows plots of two
color indices (C.I.), mag(3890\AA)-- mag(9745\AA) and mag(4210\AA) --
mag(9190\AA), versus mag(6075\AA) for all known member galaxies in the central
region of Abell 2634. For 60 early-type galaxies, there exists a clear
correlation between color and brightness in the sense that the brighter
galaxies are redder, which is known as the color-magnitude effect. This
effect is comparatively remarkable for color mag(3890\AA)$-$
mag(9745\AA) (in Fig. 4a), with a linear fit of $C.I.= -0.16 (\pm 0.03) m_h +
5.70 (\pm 0.48)$. The correlation coefficient is 0.58 and the $rms$ dispersion
is 0.17. The error on the intercept does not take into account the error on
the slope. The color index, mag(4210\AA) $-$ mag(9190\AA), also shows
such a correlation, with a linear fit of $C.I.= -0.12 (\pm 0.03)
m_h + 4.43 (\pm 0.43)$, which is shown in Fig. 4(b). Its corresponding 
correlation coefficient is 0.52 and the $rms$ dispersion is 0.15. 
The linear fits are plotted in solid lines, and the dashed lines in panel (a)
and (b) represent $\sigma_{C.I.} = 0.48$ and 0.43, respectively. 

The elliptical galaxies in the Virgo and Coma clusters are found to have a
tighter color-magnitude correlation than lenticulars do (Bower, Lucey \& Ellis
1992). However, as mentioned above, we can not accurately separate ellipticals
from lenticulars in our sample. The lenticulars in our sample might contribute
considerably on the dispersion of the color-magnitude relation.  For the sake
of comparison, the late-type galaxies and the remaining members without
morphological information are also plotted.  It is evident that late-type
galaxies do not obey such a correlation, and some galaxies with unknown 
morphologies have the color indices significantly different than the
correlation,  similar to some spirals.  


\vskip 5mm

\centerline{\framebox[10cm][c]{\Large \bf Here is Figure 4 (a)(b)}}
 \vskip 5mm

To understand the color-magnitude relation, the SEDs of three member
ellipticals with  considerably different magnitudes are shown in Figure 5. The
reddening of bright galaxies occurs mainly in the violet region shortward of
4250 \AA, corresponding to the wavelength domain of filter $b$ and $c$, where
stellar spectra are richest in metal absorption lines and the SEDs turn to be
rather steep. It has been well demonstrated that the BATC filter system has
a great advantage to study this effect. The filters $b$ and $c$ have
band widths of $\sim$ 300 \AA ~centered on 3890 \AA ~and 4210 \AA, and the
wavelengths of filters $p$ and $o$ reach longward of 9000 \AA. The
color-magnitude correlation can be used to analyze the membership possibility
of the faint objects within our observational field.

\vskip 5mm
\centerline{\framebox[10cm][c]{\Large \bf Here is Figure 5}}

\subsection{A test of the technique of photometric redshift}

It is well known that the technique of photometric redshifts can be
used to estimate the redshifts of galaxies by using the SED information from
the multicolor photometric observations. Particularly, with the development of
large and deep field surveys, this technique has been extensively applied 
(Lanzetta et al. 1996; Arnouts et al. 1999; Furusawa et
al. 2000). The photometric redshift of a given object, $z_{phot}$,
corresponds to the best fit of its photometric SED by the set of template
spectra. We also developed some procedures for estimating the photometric
redshifts, especially for the multicolor photometric observations by the
BATC filter system (Huang et al. 2001), based on the standard SED fitting
code called {\em Hyperz} (Bolzonella, Miralles \& Pell\'o 2000). It is
crucial to test the reliability in all cases by comparison between the
photometric and the spectroscopic redshifts, using the sample of relatively
bright objects.

The photometric redshift technique is traditionally utilized to search for 
galaxies or AGNs with comparatively high redshifts. The accuracy of 
photometric redshift is heavily dependent upon the photometric accuracy.  
Our sample of 124 member galaxies allows an opportunity to check our
photometric accuracy and to test whether our SED fitting procedures is
sensitive for estimating the redshifts of nearby bright galaxies. Thanks to
the large number of filters, from $b$ (close to traditional $U$ filter)
to $p$ (near-IR) bands, the accuracy of $z_{phot}$ estimates is
expected to be improved. In this calculation, we only use the SED templates of
normal galaxies, and the reddening law with a $A_V \sim 0.3$ is taken  from
Calzetti et al. (2000).  The searching step of redshifts is 0.005. 
The comparison between the $z_{phot}$ and spectroscopic redshift ($z_{sp}$) 
for 124 known member galaxies are shown in Fig. 6. The dotted line corresponds 
to $z_{phot} = z_{sp}$, and the error bar in $z_{phot}$ determination at 68\% 
confidence level is also given. 
It is obvious that the majority (98/124) of the member galaxies are
found to have $z_{phot} < 0.06$; there is a dense group at $z_{phot} \sim
0.031$, the exact redshift of Abell 2634. Some galaxies with larger
deviations are likely to have peculiar features in reddening and
metallicity evolution. The results demonstrate that our SED fitting
procedures are quite efficient in photometric redshift estimation for
galaxies belonging to such a nearby cluster. After a careful star/galaxy
separation for all objects detected, we will apply in following section the
technique of photometric redshift to the faint galaxies for selecting the
faint galaxies probably belonging to Abell 2634.

\vskip 5mm

\centerline{\framebox[10cm][c]{\Large \bf Here is Figure 6}}
 \vskip 5mm

The morphological information is also provided by the SED
fitting procedures. If we just classify member galaxies into early-types and
late-types, the classification  by SED fitting technique is quite accurate
for the early-types.  Among 60 known early-type members, only 3 early-types are
incorrectly classified as the late-types. On the other hand, the
misclassification rate for the late-type galaxies are comparatively high, 5 of
9 spirals are misclassified as early-types. So it is likely to include more
ellipticals for our SED fitting procedures. For remaining 52 known members
without morphological information in previous literatures, we shall take the
classification given by our SED fitting procedures in the following analyses.

\section{Selection of faint member galaxies of Abell 2634}
\subsection{The star/galaxy separation}


There are 5572 objects detected by our photometric measurements, and most of
them should be the faint stars. We know that the spectra of stars are
significantly different from type to type, and they are also distinct from
those of galaxies. It should be easy to separate galaxies from the whole
list of faint sources based on the SED information. 

Generally, the color-color diagram is a powerful tool because different
objects with distinct color indices usually occupy different regions in
the diagram. Figure 7 shows four types of color-color diagrams which we will
utilize in further star/galaxy separation.  These diagrams include
the following categories of objects: (i) all types of stars in our SED
template library (denoted by ``*''), (ii) morphologically various galaxies
with template SEDs (denoted by ``${\circ}$''), (iii) the known member
galaxies in  our field (denoted by ``$\times$''), and (iv) all the remaining
objects detected by our photometric measurements (denoted by dots). The most
striking feature is that the stars in our SED template library lie in a
well-defined band stretching from upper left to lower right, with which the
majority of objects detected in our observations are associated. This band
is a color sequence, with the bluer (early-type) stars at the upper left and
the redder (late-type) stars at the lower right, in the similar sense of the
main sequence in H-R diagram. However, the known galaxies are distributed in
a region just above the band, and the objects in a relatively dense region
(`bar-like') are populated by most early-type (E,S0) member galaxies. The
dashed lines in Fig. 7 can be used to set the limits between stars and
galaxies. 

\vskip 5mm

\centerline{\framebox[12cm][c]{\Large \bf Here is Figure 7 (a)(b)(c)(d)}}

The color-color diagram (a) in Fig. 7 is the best tool to
distinguish galaxies in our SED catalog. The magnitudes in the $h$ filter
(6075\AA) are available for the majority of the sources detected. The numbers
of objects detected in $b$ (3890\AA) and $p$ band (9745\AA) are relatively
small mainly because of the low intrinsic fluxes in these two bands. Certain
number of objects in our SED list are found to have not been detected in $b$
or/and $p$ bands, for which we can't use diagram Fig. 7(a) to perform
star/galaxy separation. Alternatively, the magnitudes in $c$ (4210\AA) and
$o$ (9190\AA) bands can be used for star/galaxy separation. 

In practice, we excluded all the known galaxies from the list of 5573 objects,
and then divided the remaining objects into following four categories: 
(i) 1157 objects with $b$,$h$, and $p$ magnitudes; 
(ii) 122 objects with $b$,$h$,and $o$ magnitudes, but without $p$ magnitude; 
(iii) 401 objects with $c$,$h$, and $p$ magnitudes, but without $b$ magnitude; 
(iv) 220 objects with $c$,$h$, and $o$ magnitudes, but without $b$ and $p$
magnitudes. It is appropriate for these four categories to produce
the star/galaxy separations by those four color-color diagrams 
given in Fig. 7. As a result, we obtained 359 faint galaxies in our field;
the technique of photometric redshift will be applied for membership
selection to those in the following section.

\subsection{The membership selection by photometric redshift estimation and
color-magnitude correlation}

The results of the application of photometric redshift
technique on 124 known member galaxies opened a straight way for
membership selection. To produce the list of faint member galaxies, we
applied the technique of photometric redshift upon 359 faint galaxy
candidates. We chose the same parameters in the testing
application except taking a smaller step of photometric redshift, 0.002
(i.e., 600 km s$^{-1}$).  The histogram of estimated photometric redshifts
are shown in Fig. 8. Most of candidates are found to have $z_{phot}$
less than 0.25. Certain fraction of galaxies lie in the bump within a region of
$z_{phot}$ from 0.0 to 0.06, corresponding to a radial velocity  limit $V_r <
18000$ km s$^{-1}$. About 80\% known galaxies were computed to have the
photometric redshifts within this $z_{phot}$ region (see Fig. 6b). Therefore,
we adopt 0.06 as the limit between members and non-members, and 76 galaxies
are selected to be member candidates.  

\vskip 5mm

\centerline{\framebox[12cm][c]{\Large \bf Here is Figure 8}}

Among the 76 member candidates, 68 (90\%) galaxies are detected as early-type
galaxies by our SED fitting procedures, including 60 E and 8 S0 galaxies. 
The testing application on 124 known members shows that some late-type
galaxies, particularly for Sa galaxies, are likely to be misclassified as
early-types. For these 68 early-type member candidates, a further selection by
color-magnitude correlation is performed. Figure 9 gives the color-magnitude 
diagram for all 359 galaxy candidates, including 68 newly-selected member 
candidates (denoted by ``$\odot$'').  We find that a majority of early-type 
member candidates obey the previous form of color-magnitude relation 
that is derived only by bright early-type galaxies.  There are 10 
candidates with colors beyond the $2\sigma$ deviation of intercept which 
have probably been misclassified. Taking only the templates of
late-type galaxies, we applied the SED fitting procedures again on these 
10 member candidates. As a result, 8 of 10 galaxies are found to have 
redshifts less than 0.06, with best fit to the template SEDs of Sa or Sb
galaxies, and they are therefore regarded as late-type member galaxies. The
remaining 2 galaxies with larger photometric redshifts are excluded.

Finally, 74 galaxies (including 58 early-types and 16 spirals) are selected
as faint member galaxies in the central region of Abell 2634. The SED
information for these 74 faint members is cataloged in Table 3, as well as
their positions, $z_{phot}$ values, and morphological classes $T$ (E, S0,
Sa, Sb galaxies are represented by 1, 2, 3, 4, respectively).  

\vskip 5mm
\centerline{\framebox[12cm][c]{\Large \bf Here is Figure 9}}  


\section{Spatial and color properties of our enlarged sample}

Combining with the 124 known member galaxies, we achieved a larger sample of
members in the central region of Abell 2634.  Statistically, there are 198
member galaxies, including 160 early-types (102 E, 56 S0, 2 S0a), 37
late-types (13 Sa, 18 Sb, 6 Sc), taking the morphological classification given
by our SED fitting procedures for 52 known members without morphological
information in the literature.

Based on the enlarged sample of member galaxies, we can observe the spatial
distribution in the central region of Abell 2634. Fig. 10 gives the spatial
distribution of all member galaxies, with respect to the central galaxy
NGC 7720.  Different classes of galaxies are denoted by different symbols. 
It is easy to see that the early-type galaxies  dominate in the central region, 
exactly superposing on the position of NGC 7720. The contour map of the 
surface density for all early-type galaxies, using a 5 $\times$ 5 $arcmin^2$
smoothing window, are superposed. The contour levels in Fig.10 are 0.03, 0.08, 
0.13 and 0.18 $arcmin^{-2}$, respectively. However, the spatial property of 
late-type galaxies is
significantly different from that of the early-types. The distribution of
spirals is rather scattered, and likely to locate in the outer region of Abell
2634, to the south-east of NGC 7720.

\vskip 5mm
\centerline{\framebox[12cm][c]{\Large \bf Here is Figure 10}} 
\vskip 5mm


The color-magnitude relation of early-type galaxies has commonly been
interpreted as a stellar metallicity effect  in the sense that mean stellar
metallicities become progressively higher in brighter galaxies (Arimoto \&
Yoshii 1987). However, this point of view was recently challenged by a stellar
age effect , claimed by Worthey (1996), that the observed color-magnitude
relation may result from the younger mean stellar ages in fainter galaxies. 
Furthermore, Dressler (1980b) found a well defined relationship between local 
galaxy density and galaxy type, indicating an increasing elliptical and
lenticular population and a corresponding decrease in spirals with increasing
density. Our enlarged sample of member galaxies in central region of
Abell 2634 includes 160 early-type galaxies and 36 spirals, which allows an
illustration on relations between color and magnitude, and between local
density and galaxy type.

Fig. 11(a) shows the color-magnitude relation for all members in the central
region of  Abell 2634; the relation between color and angular separation, 
related to NGC 7720 for all members, is given in panel (b). The color is
represented by the $C.I.$, mag(4210\AA) -- mag(9190\AA). The faint blue
galaxies, covering the lower right region of panel (a) with $C.I. < 2.0$ and
$h$ mag $>$ 16.5, are dominated by the late-type galaxies. Most of blue
galaxies are found to be scattered in the outer region of this cluster, with
an angular separation more than 10 arcmin. On the other hand, The dense
core of Abell 2634, defined by an angular separation less than 10 arcmin, is
populated by early-type galaxies, which is also well shown in Fig. 10. Our
results are consistent with the morphology-density relation found by 
Dressler (1980b).

It should be stressed that there exists a tendency that the dispersion of color
indices of the early-type galaxies becomes larger in the outer region (see
panel (b)). This phenomenon is first reported in this paper; it might
reflect some clues about the {\em environmental} effect on the evolution of
galaxies in a cluster. Assuming that the larger color indices are due to the
higher mean stellar metallicities in early-type galaxies, the denser
environment in the innermost region is likely to enhance the physical 
processes which lead to a increase of mean stellar metallicities. The
scenario of spiral-spiral merge provides a sound interpretation of the 
scarcity of late-type galaxies in the core region.  We also notice that
the reliability of this phenomenon depends on the combined sample.  As
shown in Fig. 9, the galaxies fainter than 16.0 mag are dominant in the 
list of newly-selected members, and they are more likely to have lower color 
indices, which contributes a lot to the dispersion of color indices.  
Therefore, it is important to verify this result by taking the follow-up 
spectroscopic observations of these new member candidates.

\vskip 5mm

\centerline{\framebox[12cm][c]{\Large \bf Here is Figure 11(a)(b)}} 
%

\section{Summary}

This paper presents our multicolor optical photometry for the central region
($56' \times 56'$) of Abell 2634, using the 60/90 cm Schmidt Telescope of
Beijing Astronomical Observatory equipped with 14 BATC filters which cover
almost the whole optical wavelength domain. As a result, we obtained the SEDs
of 5572 objects. With the help of NASA/IPAC Extragalactic Database and the
compilations in the literature, we selected 124 known member galaxies from
our SED lists, for which the detailed analyses on their SED features are
performed. The early-type galaxies are dominant in the central region of
Abell 2634. The sample of 59 known early-types offers an opportunity to
check the accuracy and reliability of our SED data by the comparison between
the observed SEDs and template SEDs. We applied the SED fitting technique
on 124 known member galaxies to estimate the photometric redshifts, and
obtained good results: 80\% of the members have the $z_{phot}$ less
than 0.06, which provides a reliable tool for further membership selection.
Furthermore, a verification of color-magnitude correlation using our SED
information of early-types demonstrates the great advantages and potential
for BATC multicolor observations in studying the color-magnitude effect. 

Based on the knowledge of SED features of known member galaxies, we carefully
utilized the color-color diagrams, technique of photometric redshift, and
color-magnitude correlation (particularly for early-type members) in
the selection of faint members. As a result, we isolated 74 galaxies as the
faint members, including 58 early-types and 16 spirals. Then, based on the
enlarged sample of member galaxies, the spatial distributions for
morphologically various galaxies are discussed. We found that most of blue
galaxies are scattered in  the outer region of this cluster, and core region
is populated by early-type  galaxies for Abell 2634. Furthermore, the color
dispersion for early-type galaxies is larger in the outer region, which might
be a kind of environmental effect on the evolution of cluster galaxies.

Based on our SED catalogs of known and new members, some studies on the
evolution of galaxy populations in the Abell 2634 can be carried out.
In order to investigate the dynamics and kinematics of the galaxies in the
central region, the follow-up efforts of spectroscopic observations for
sake of obtaining the accurate radial velocities for all the faint member
galaxies can be proposed. 

\vskip 1cm

\acknowledgments

This research has made use of the NASA/IPAC Extragalactic Database (NED) 
which is operated by the Jet Propulsion Laboratory, California Institute of
Technology, under contract with the National Aeronautics and Space
Administration.  This work is mainly supported by the National Key Base
Sciences Research Foundation under the contract G1999075402, is also
partly supported by the Chinese National Science Foundation (NSF) under the
contract No.19873007 and 19873018 to Yuan, Q., NSF No.19833020 and 19503003
to X. Zhou. We would like to thank Mr. Yaohua Li for his management and
support of the instruments, and Ray Chen, Haotong Zhang, Bing Zhao, Zheng
Zheng for obtaining the observational data and sharing their experience with
us. We also appreciate the assistants who contributed their hard work  to the
observations. Furthermore, Yuan wants to express his sincere gratitude to Dr.
Richard Green for the hospitality during his visit to National Optical
Astronomy Observatories where the first version of this article is finished,
and  Dr. Hulmut Abt for the help in improving the English writing of this
paper.

\newpage 
\def\baselinestretch{1.0}  
{\small
\begin{table}[ht]
\caption[]{The parameters of the BATC filters and the statistics of
our observations}
\vspace {0.5cm}
\begin{tabular}{cccccccccc}   \hline   \hline
\noalign{\smallskip}  
 No. & Filter & Central $\lambda$ & FWHM &
Exposure &  Number of & Seeing$^a$ & Calibration & Stars & Limiting \\
   &  name & $(\AA)$ & $(\AA)$ & (second)& images & (arcsec) & images &
detected & mag.\\
\noalign{\smallskip}   \hline \noalign{\smallskip} 
  1  & b  & 3890 & 291 & 23100  & 29    &  5.36 & 3 & 1427 & 19.5 \\
  2  & c  & 4210 & 309 &  3600  & 3      &  6.12 & 0$^b$  & 2190 & 20.0 \\
  3  & d  & 4550 & 332 & 4500  & 6       &  5.29 & 3  & 3050 & 20.5 \\
  4  & e  & 4920 & 374 & 13980  & 15   &  3.94 & 3  & 3561 & 20.5 \\
  5  & f  & 5270 & 344 & 6540  & 16     &  4.70 & 3  & 3010 & 20.0 \\
  6  & g  & 5795 & 289 & 8400  &  7     &  5.90 & 4  & 3837 & 20.0 \\
  7  & h  & 6075 & 308 & 5340  & 14    &  4.09 & 6  & 3919 & 19.5 \\
  8  & i  & 6660 & 491 & 6320  & 12     &  5.02 & 6  & 3910 & 19.5 \\
  9 & j  & 7050 & 238 & 4210  & 16      & 4.21 & 4  & 4182 & 19.0 \\
  10 & k  & 7490 & 192 & 8880  & 10   &  3.77 & 4  & 5572 & 20.0 \\
  11 & m  & 8020 & 255 & 10960  & 13 &  4.46 & 3  & 4356 & 19.0 \\
  12 & n  & 8480 & 167 & 18480  & 16  &  4.80 & 1  & 3948 & 19.0 \\
  13 & o  & 9190 & 247 &  4740  & 6     &  3.78 & 0$^b$  & 3432 & 18.5 \\
  14 & p  & 9745 & 275 & 8340  & 10    &  4.79 & 3  & 2220 & 17.5 \\
\noalign{\smallskip}   \hline  
\end{tabular}
\end{table}
\vskip -3mm   
\noindent $^a$  This column lists the seeings of
the combined images.

\noindent $^b$  No calibration image was taken. The intrinsic
magnitudes for the $c$ and $o$ filters are determined by \\
~~~the method of SED model (Zhou et al. 1999). 
}

		    
\def\baselinestretch{1.4}  

\newpage 
\thispagestyle{empty}
\begin{figure}[tt]		    
\vskip -4cm
\centerline{\psfig{file=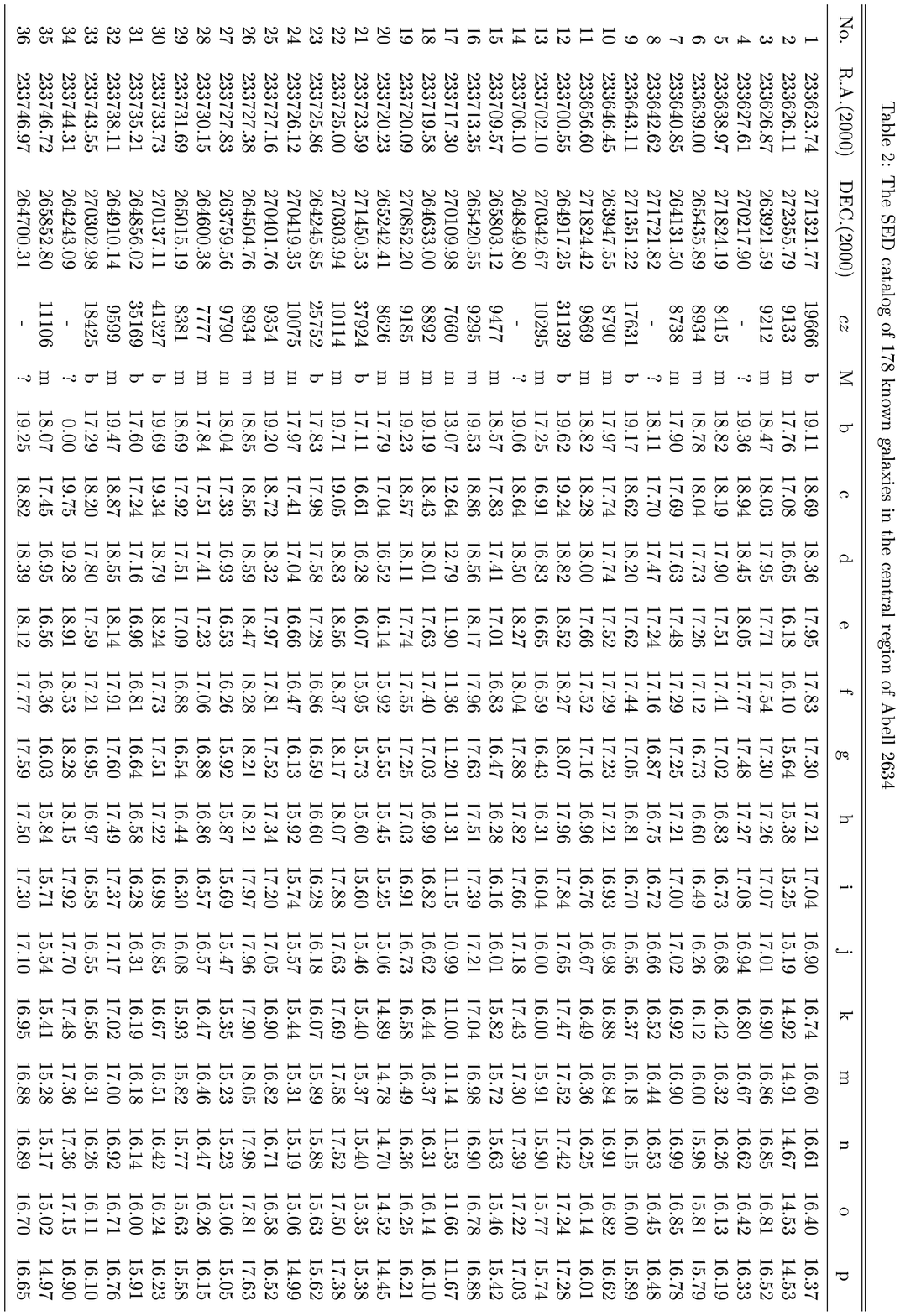,width=21cm,angle=0}}
\end{figure}			    

\newpage \thispagestyle{empty}				    
\begin{figure}[tt]		    
\vskip -4cm
\centerline{\psfig{file=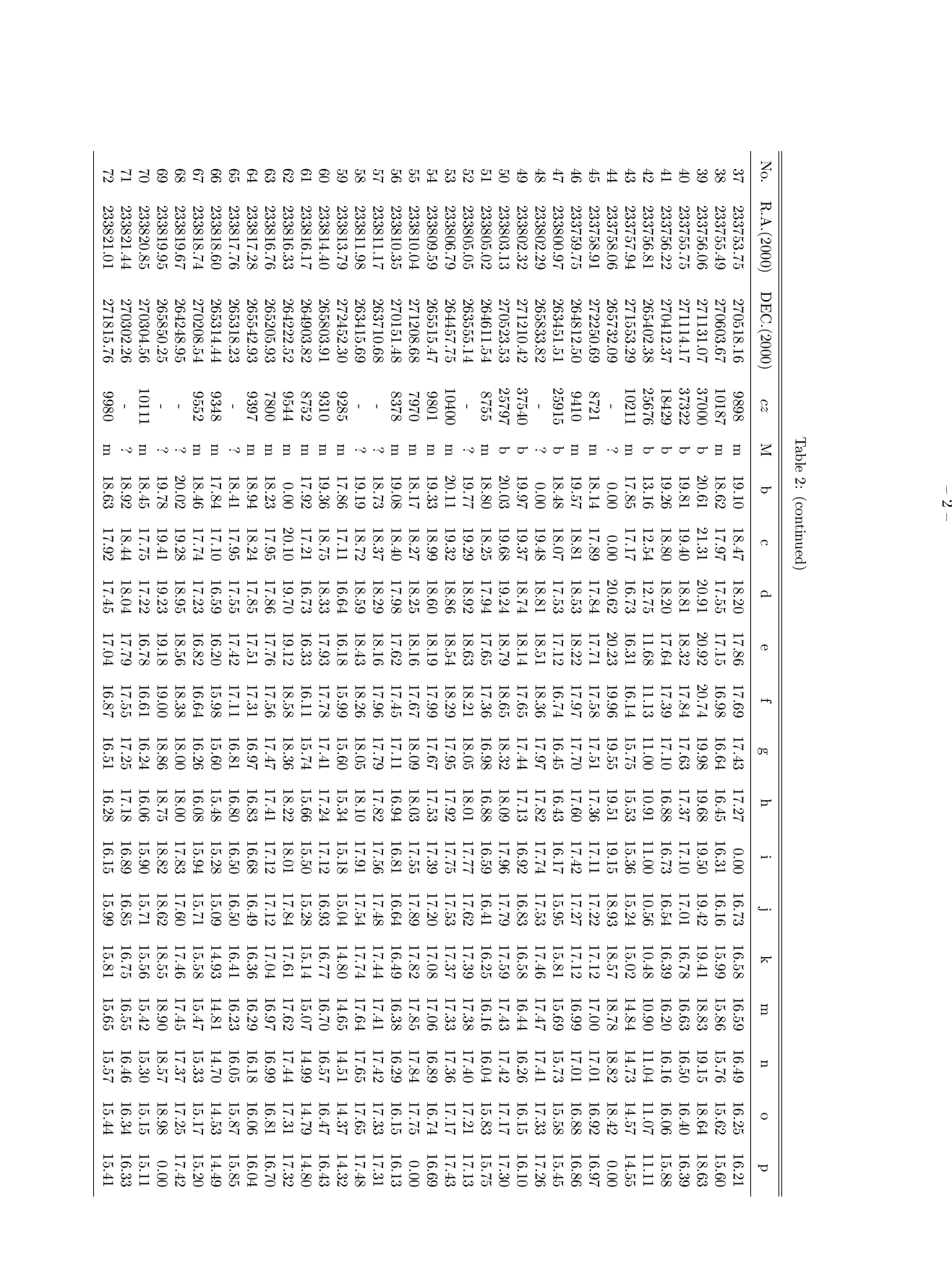,width=20cm,angle=0}}
\end{figure}			    
				    
\newpage \thispagestyle{empty}				    
\begin{figure}[tt]		    
\vskip -4cm
\centerline{\psfig{file=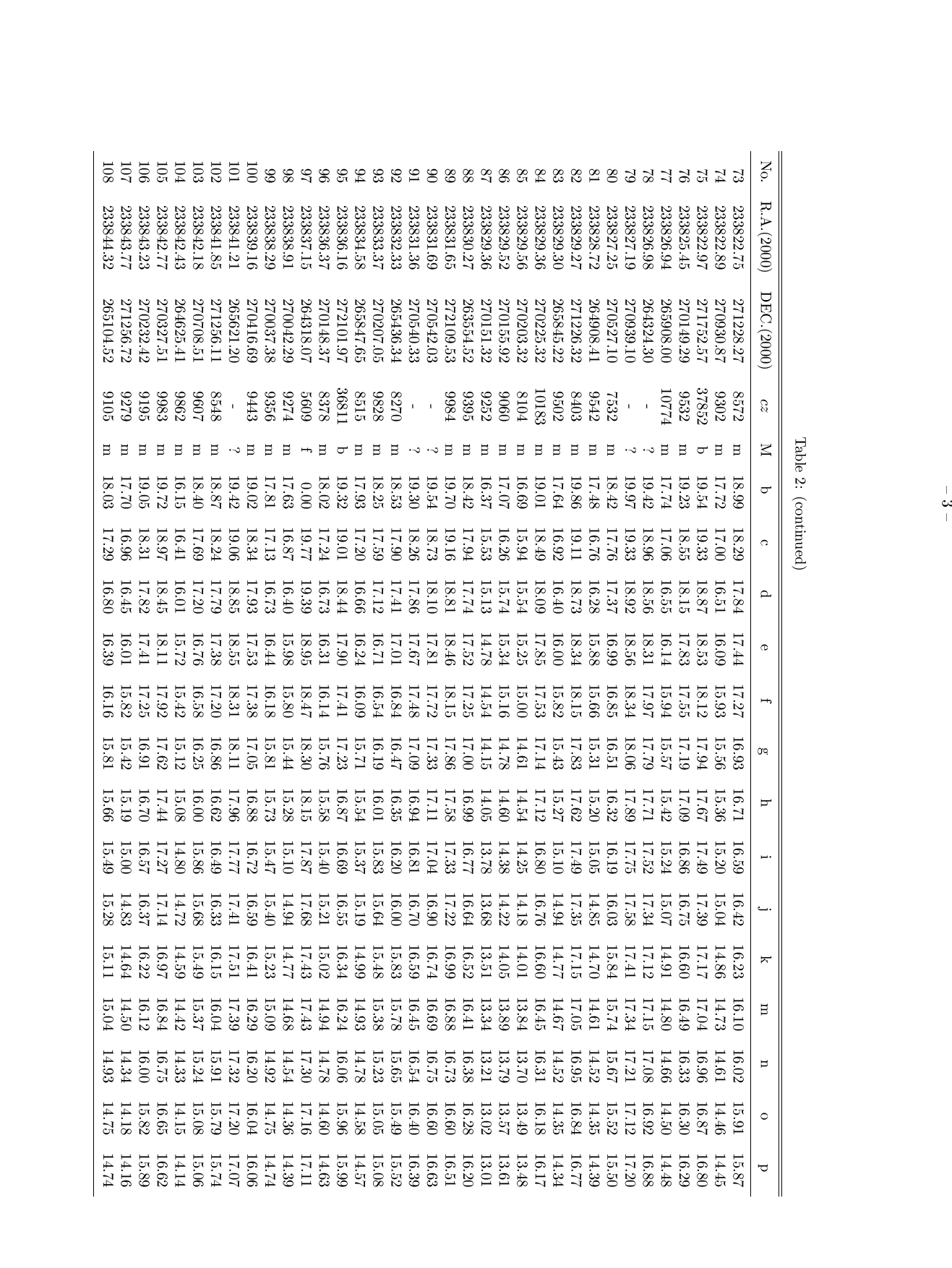,width=20cm,angle=0}}
\end{figure}			    

\newpage \thispagestyle{empty}				    
\begin{figure}[tt]		    
\vskip -4cm
\centerline{\psfig{file=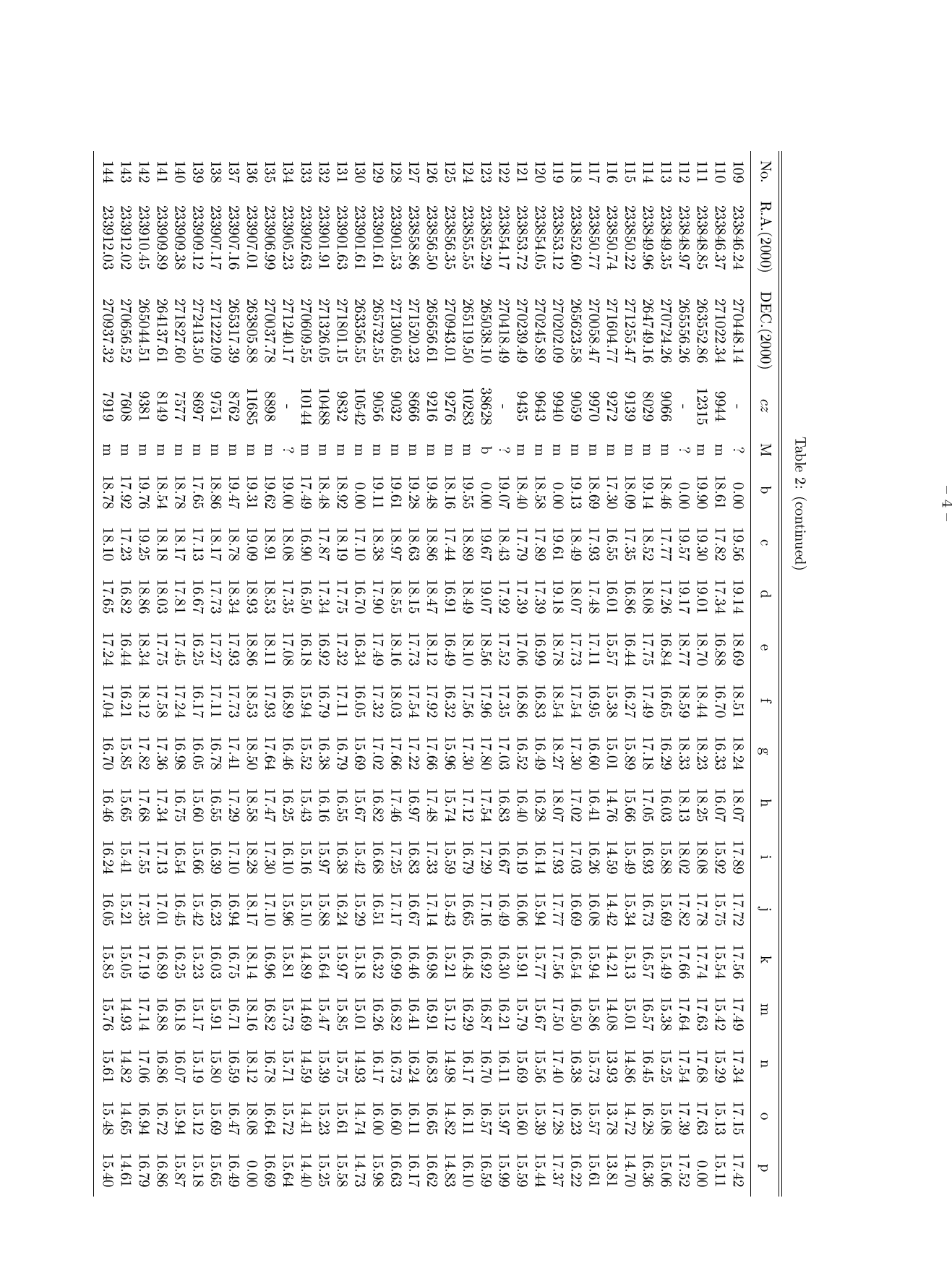,width=20cm,angle=0}}
\end{figure}			    

\newpage \thispagestyle{empty}				    
\begin{figure}[tt]		    
\vskip -4cm
\centerline{\psfig{file=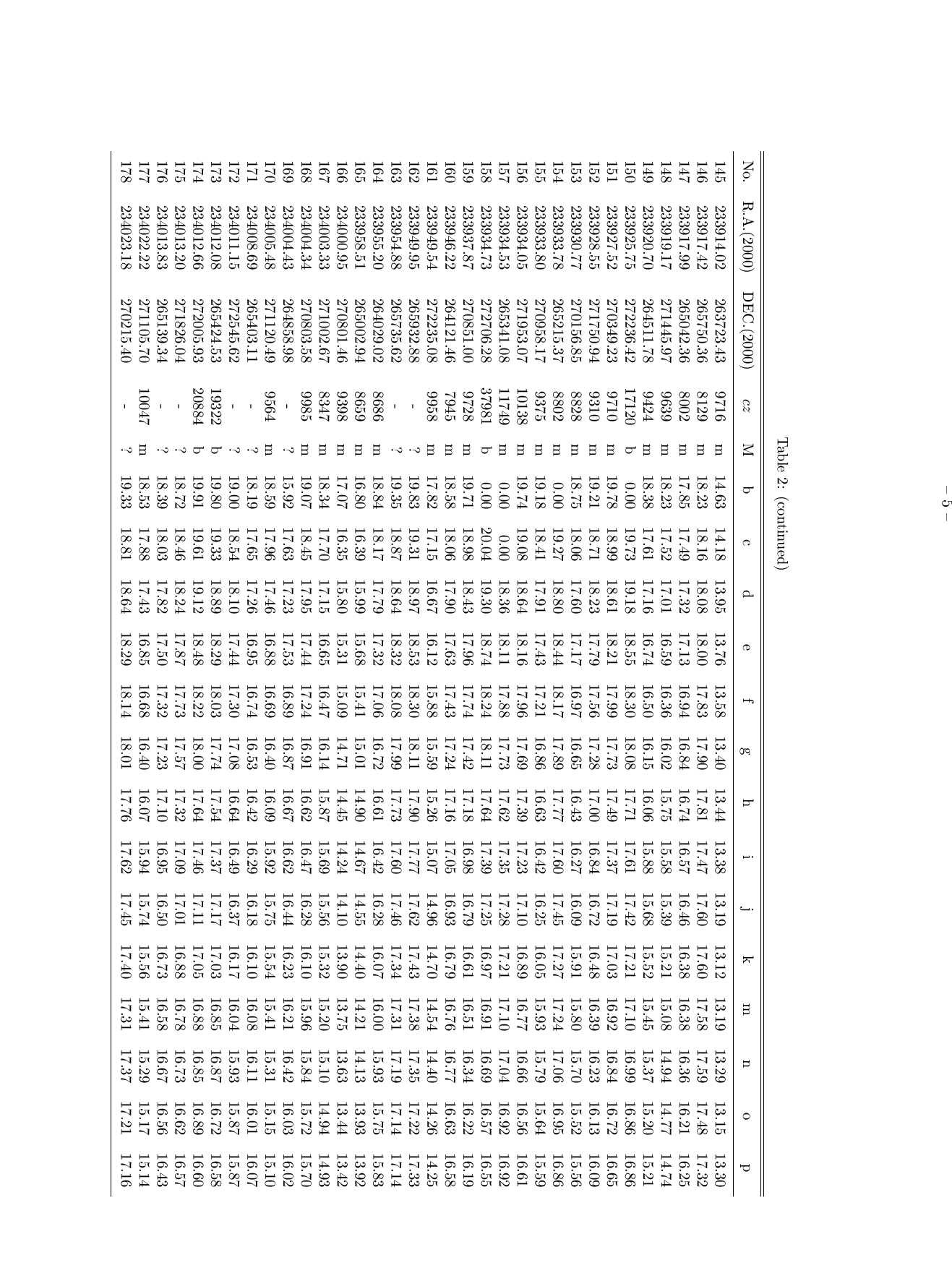,width=20cm,angle=0}}
\end{figure}

\newpage \thispagestyle{empty}				    
\begin{figure}[tt]		    
\vskip -4cm
\centerline{\psfig{file=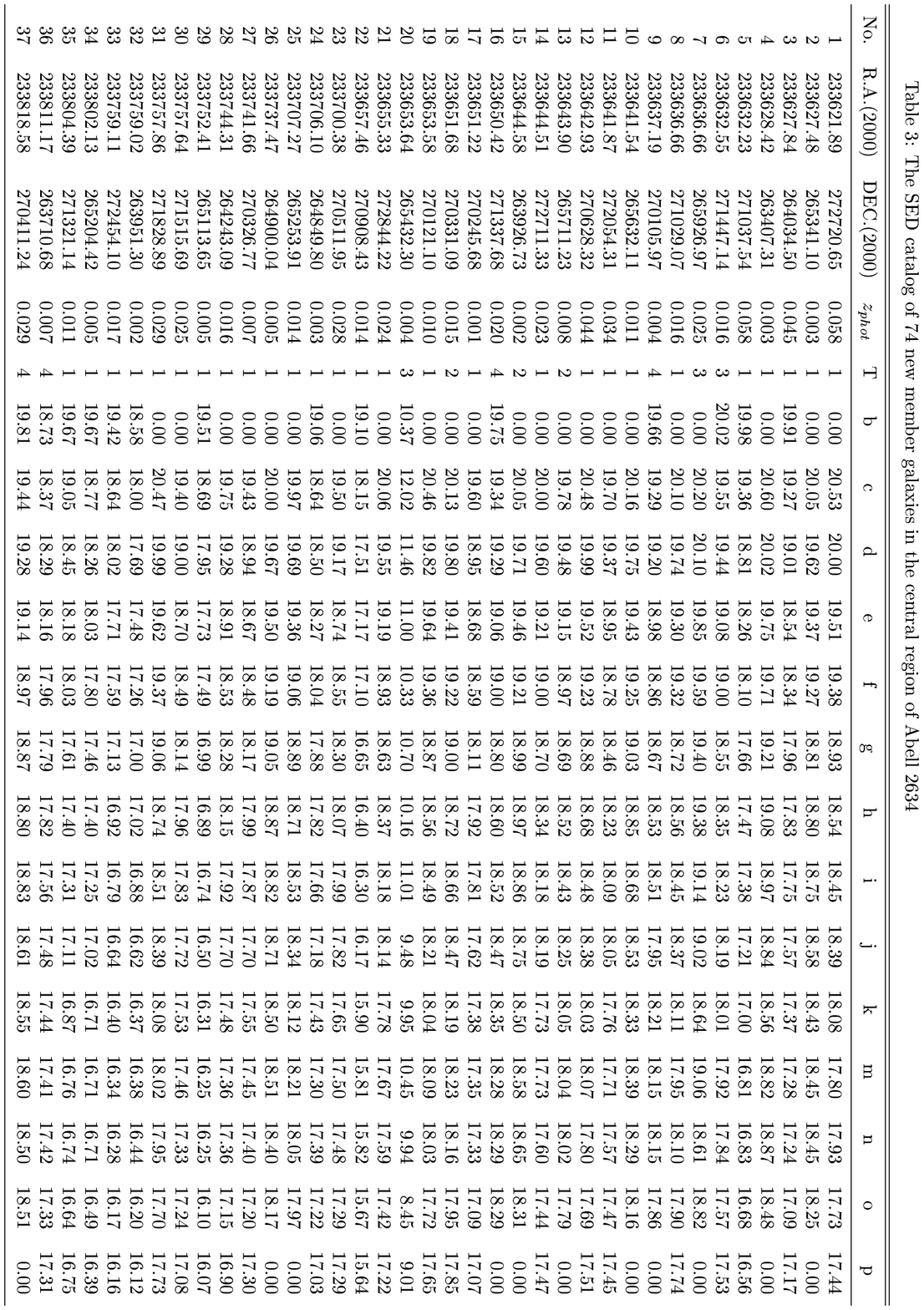,width=20cm,angle=0}}
\end{figure}

\newpage \thispagestyle{empty}				    
\begin{figure}		    
\vskip -4cm
\centerline{\psfig{file=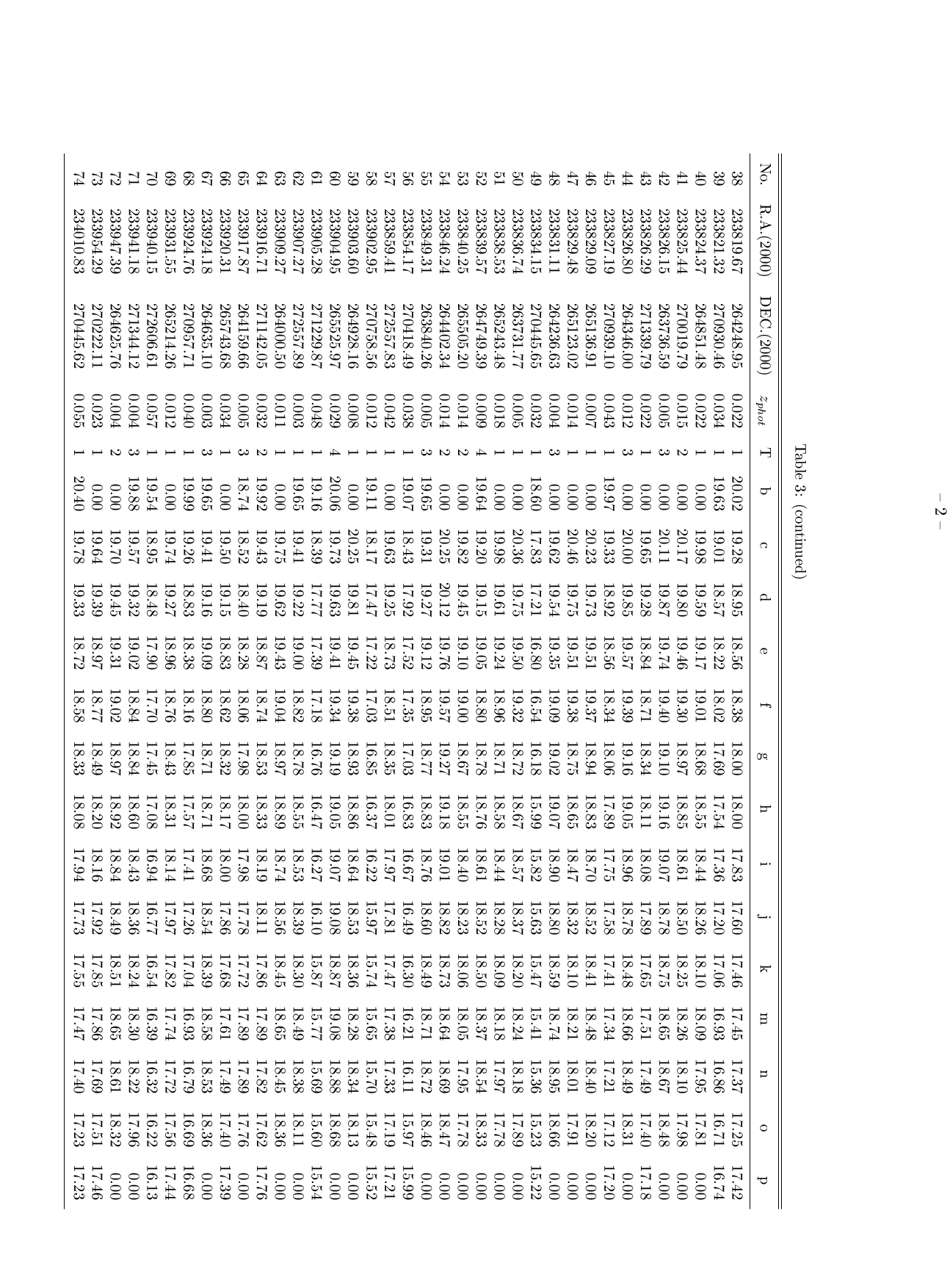,width=20cm,angle=0}}
\end{figure}

\newpage 

\begin{figure}[h]
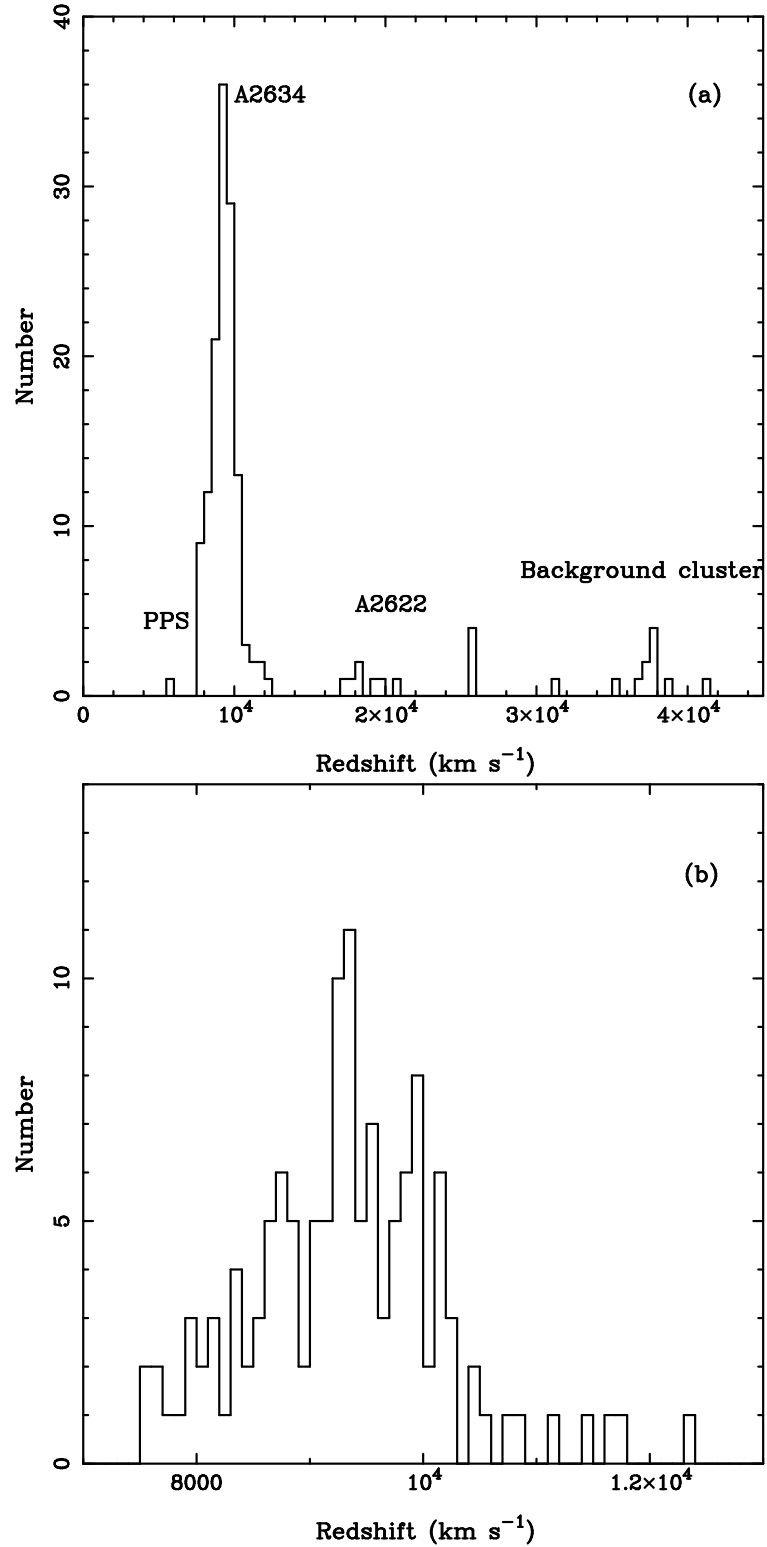

  \centerline{\psfig{figure=fig1a.ps,angle=270,width=10.0cm}}
  \centerline{\psfig{figure=fig1b.ps,angle=270,width=10.0cm}}
\caption{ Histograms of redshifts, $cz$, of known galaxies in our field.  
Panel (a) for 147 known galaxies, and panel (b) for 124 member galaxies. The
bin widths are 500 km s$^{-1}$ and 100 km s$^{-1}$, respectively. }
  \label{Fig1(a)(b)}
\end{figure}

\newpage 
\begin{figure}[h]
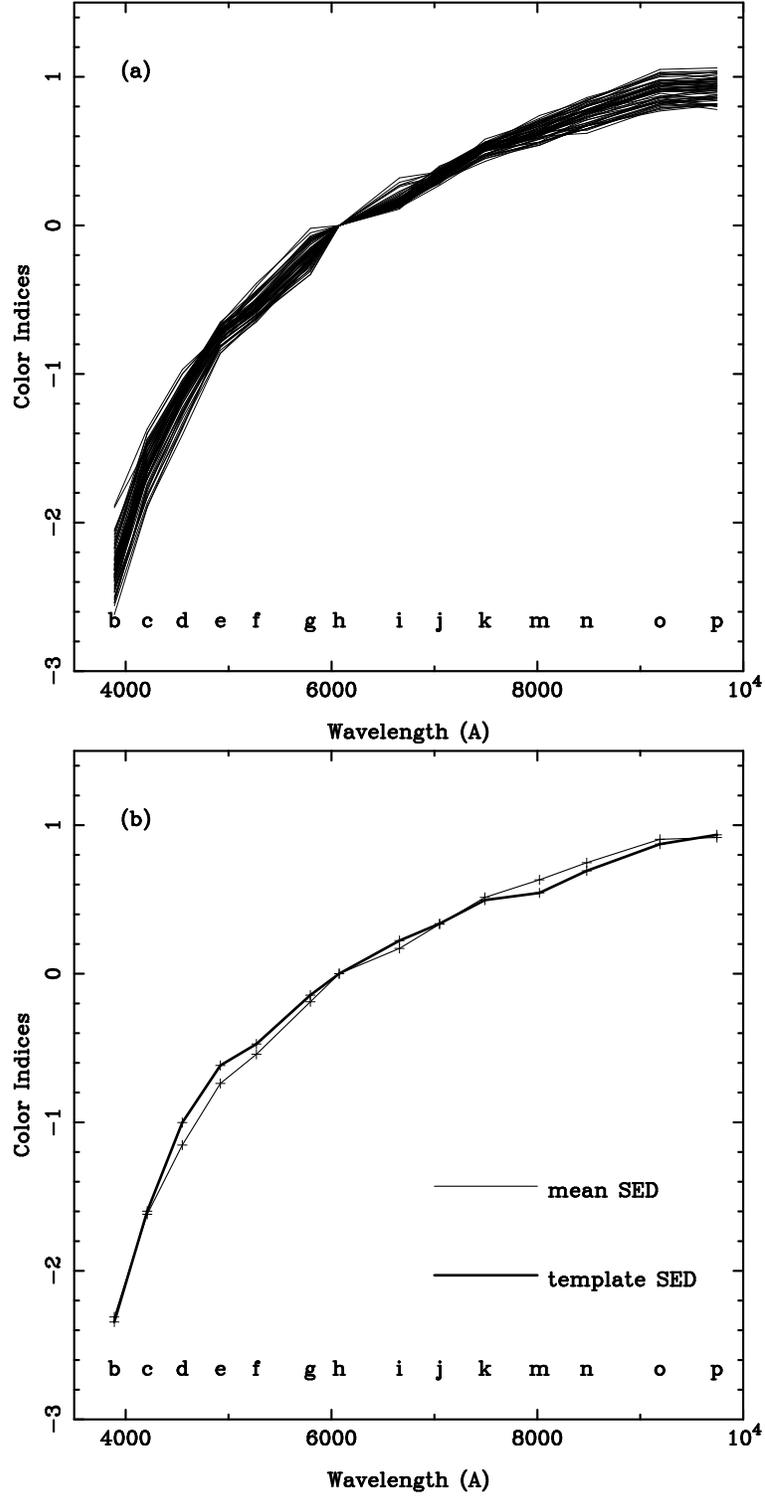

 \centerline{\psfig{figure=fig2a.ps,angle=270,width=10.0cm}}
 \centerline{\psfig{figure=fig2b.ps,angle=270,width=10.0cm}}
\caption{ The SEDs of 59 early-type galaxies in panel (a); and the mean SED
and the template SED is plotted in panel (b) with thin and thick lines,
respectively.  
 }
  \label{Fig2(a)(b)}
\end{figure}

\newpage
\begin{figure}[h]
 \centerline{\psfig{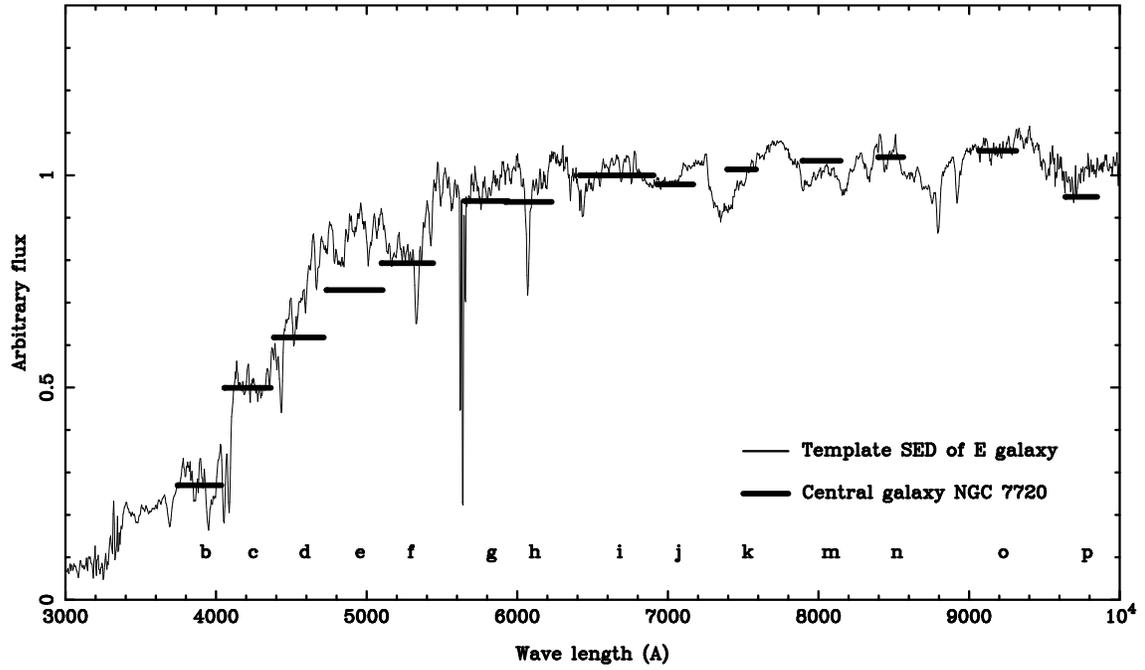}}
\caption{ The SED of the central galaxy, NGC 7720,
and the template spectrum of the elliptical galaxy.  
 }
  \label{Fig3}
\end{figure}

\newpage
\begin{figure}[h]
  \centerline{\psfig{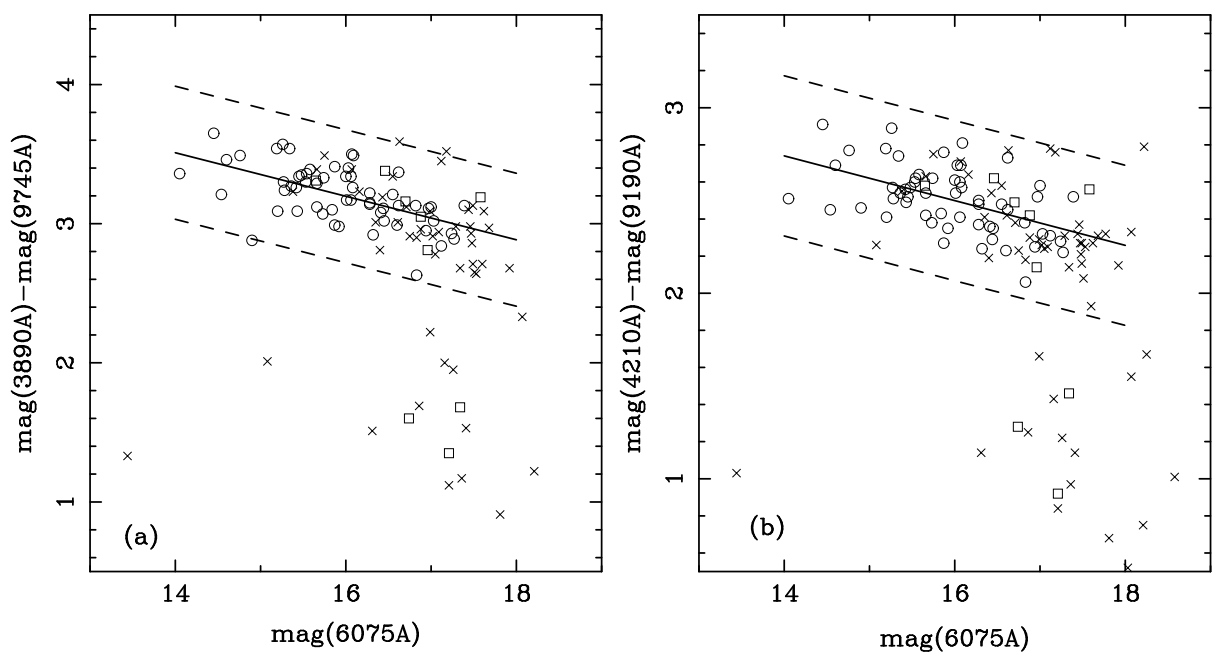}}
\caption{ The color-magnitude effect for early-type galaxies (denoted by
``$\circ$'') in the central region of Abell 2634. The color-magnitude relation
for color index  mag(3890\AA)$-$ mag(9745\AA) in panel (a), and that for color
index mag(4210\AA) $-$ mag(9190\AA) in panel (b). The colors of the known
spirals and the galaxies with unknown morphologies are denoted by``${\Box}$''
and ``${\times}$''.  The linear fits are plotted in solid lines, and the 
dashed lines in panel (a) and (b) represent $\sigma_{C.I.} = 0.48$ and 
0.43, respectively.}
\label{Fig4(a)(b)}  
\end{figure} 

\newpage
\begin{figure}[h]
  \centerline{\psfig{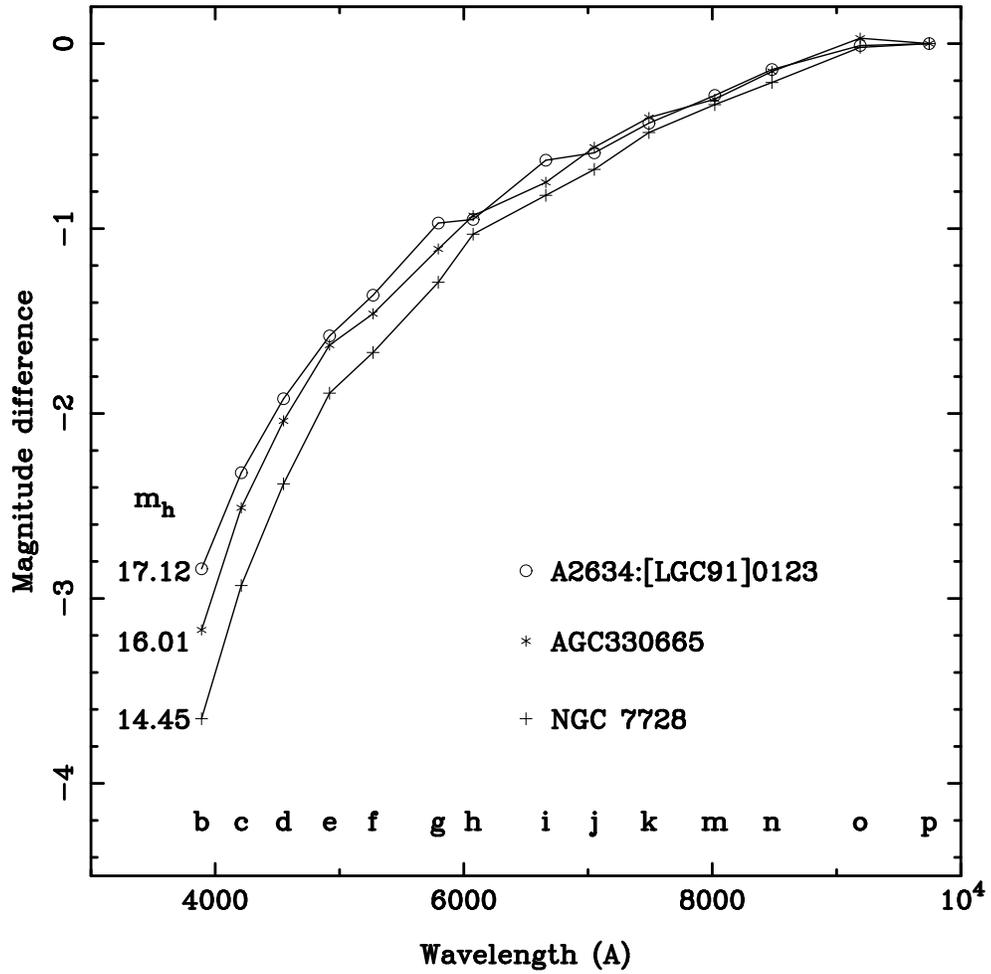}}
\caption{ The SEDs of three member ellipticals with considerably different 
magnitudes. The $h$ magnitudes and names of these ellipticals are also given.
}
 \label{Fig.5} 
 \end{figure} 

\newpage
\begin{figure}[h]
  \centerline{\psfig{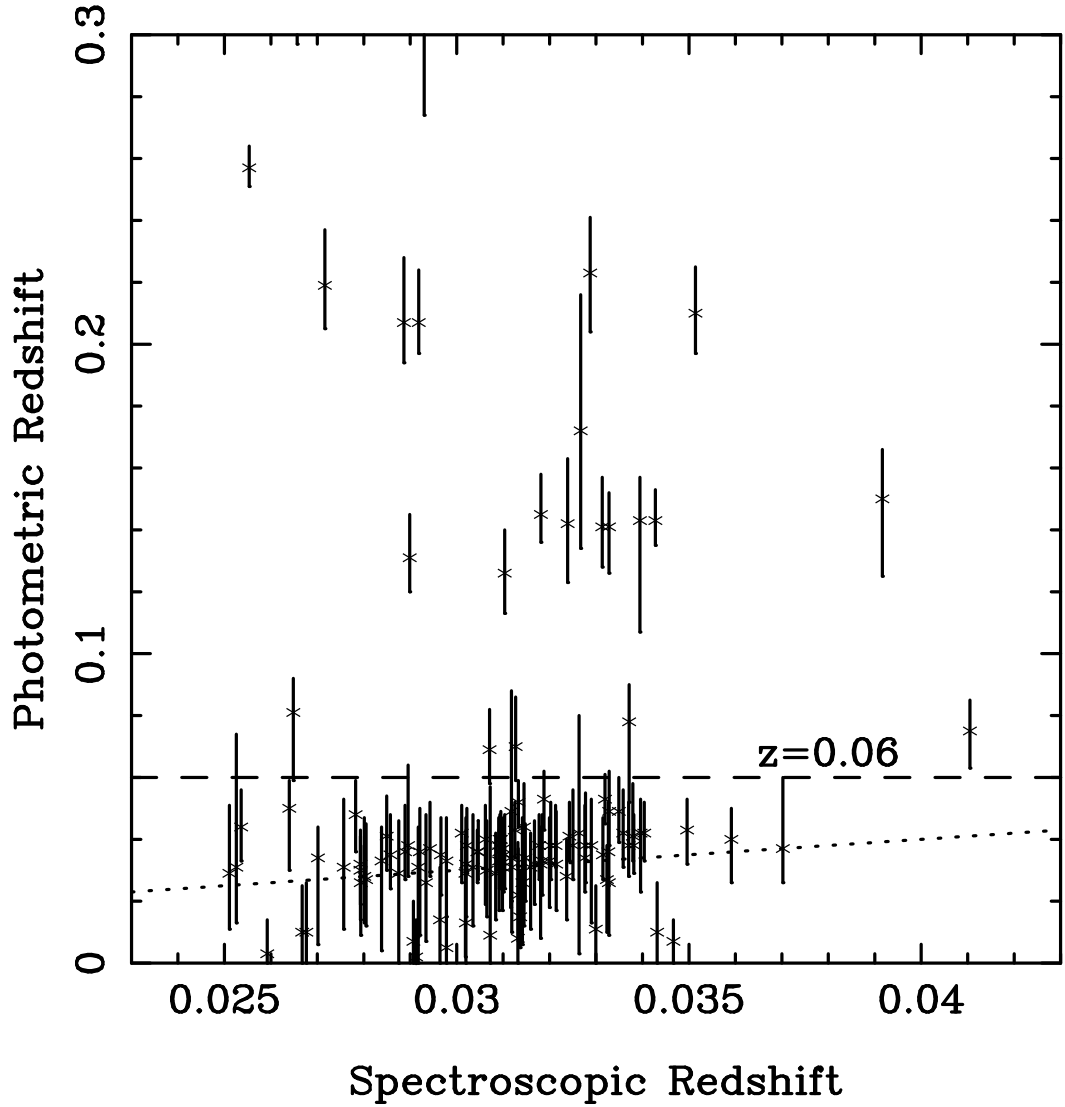}}
\caption{ 
Comparison between the photometric redshift ($z_{phot}$) and spectroscopic 
redshift ($z_{sp}$) for 124 known member galaxies. The dotted line corresponds 
to $z_{phot} = z_{sp}$, and the error bar in $z_{phot}$ determination at 68\% 
confidence level is also given. 
}   
\label{Fig6}
\end{figure}
  
\newpage
\begin{figure}[h]
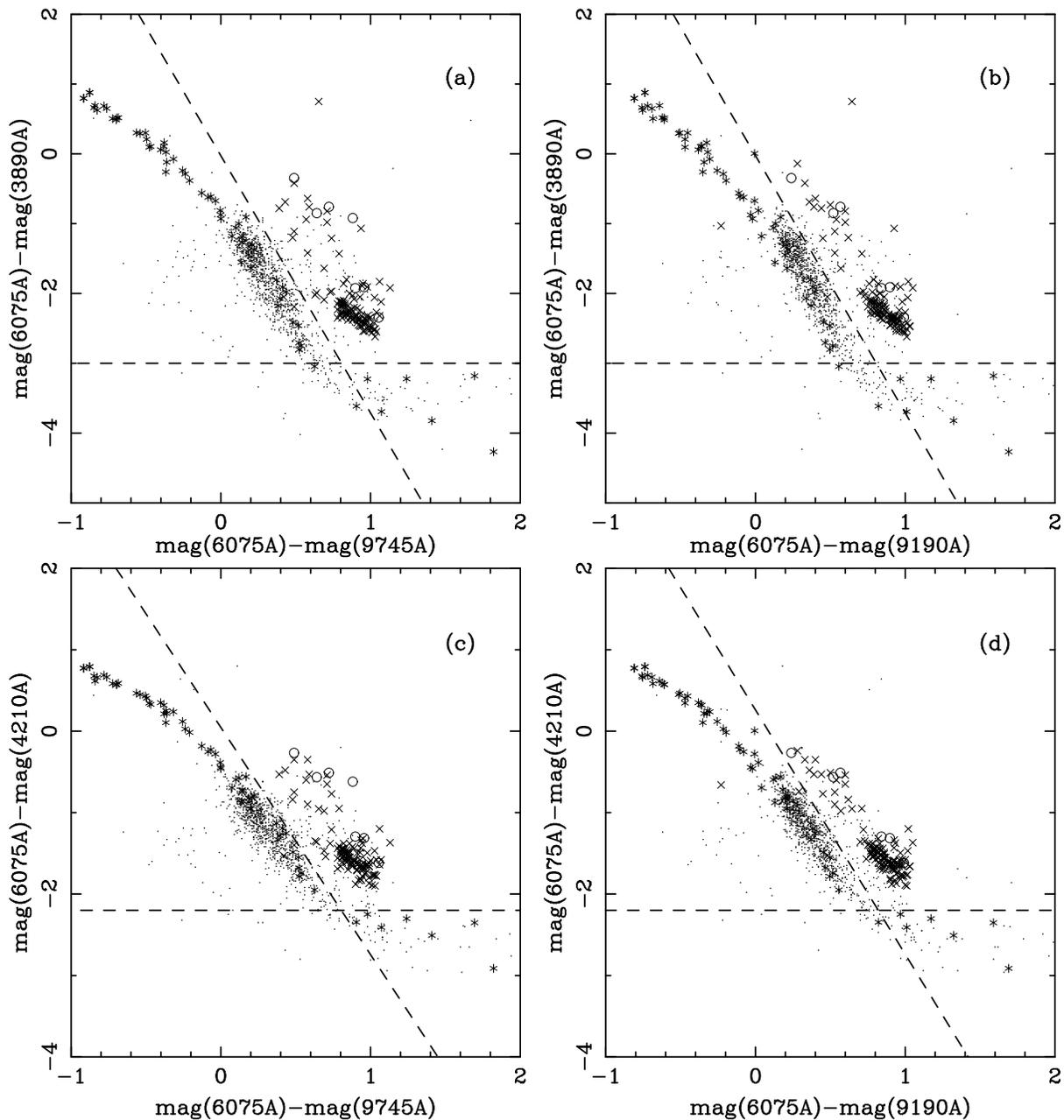

  \centerline{\psfig{figure=fig7ab.ps,angle=270,width=16.0cm}}
  \centerline{\psfig{figure=fig7cd.ps,angle=270,width=16.0cm}}
\caption{  Four kinds of color-Color diagrams used in our star/galaxy 
  separation. The color indices of symbols in the diagrams distinguish the
galaxies as follows:  ``*'' for all types of stars in our SED template library;
``${\circ}$'' for morphologically different galaxies with template SEDs;
``$\times$'' for the known member galaxies in our field; and the minor dots for
all the objects detected in $b/c$ and $o/p$ bands by our photometric
measurements.  }   
\label{Fig7(ab)(cd)}
\end{figure}

\newpage
\begin{figure}[h]
  \centerline{\psfig{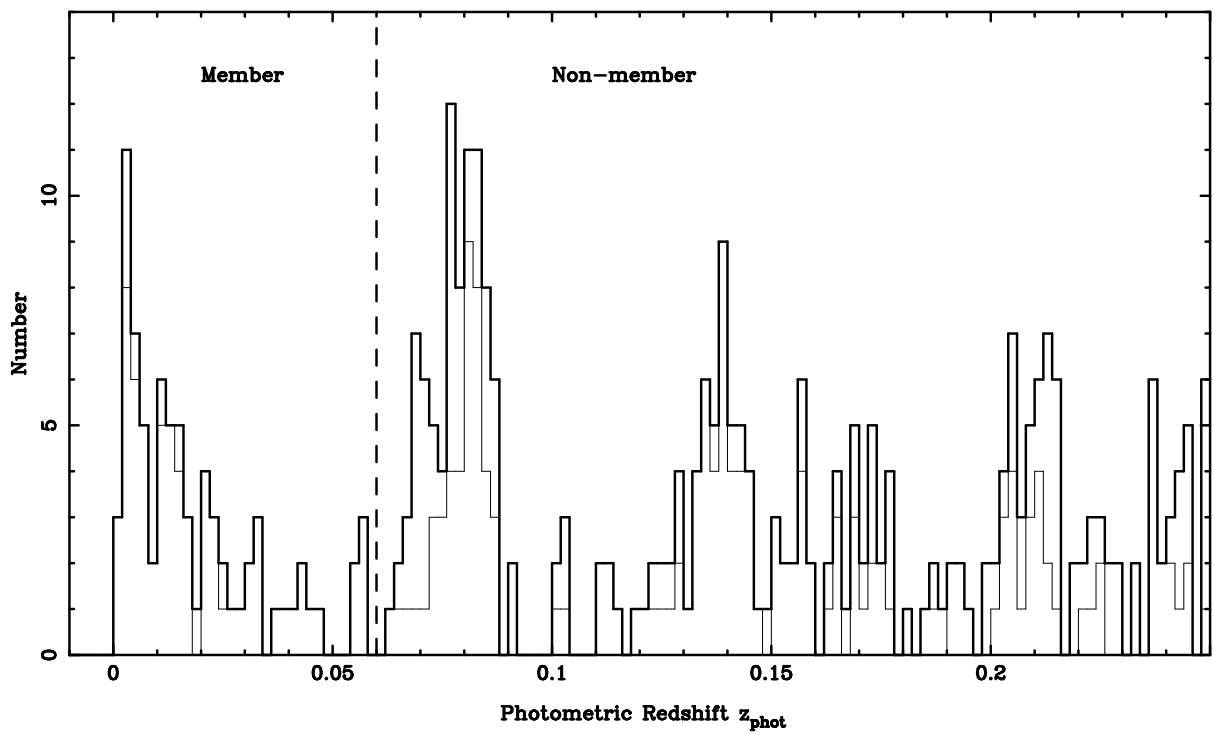}}
\caption{  Distribution of photometric redshifts for 359 faint galaxies (thick
line). The redshift of 0.06 is adopted as the limit between members and
non-members. The thin line corresponds to $z_{phot}$ distribution of
early-type galaxies. }    \label{Fig8)}  \end{figure}

\newpage
\begin{figure}[h]
  \centerline{\psfig{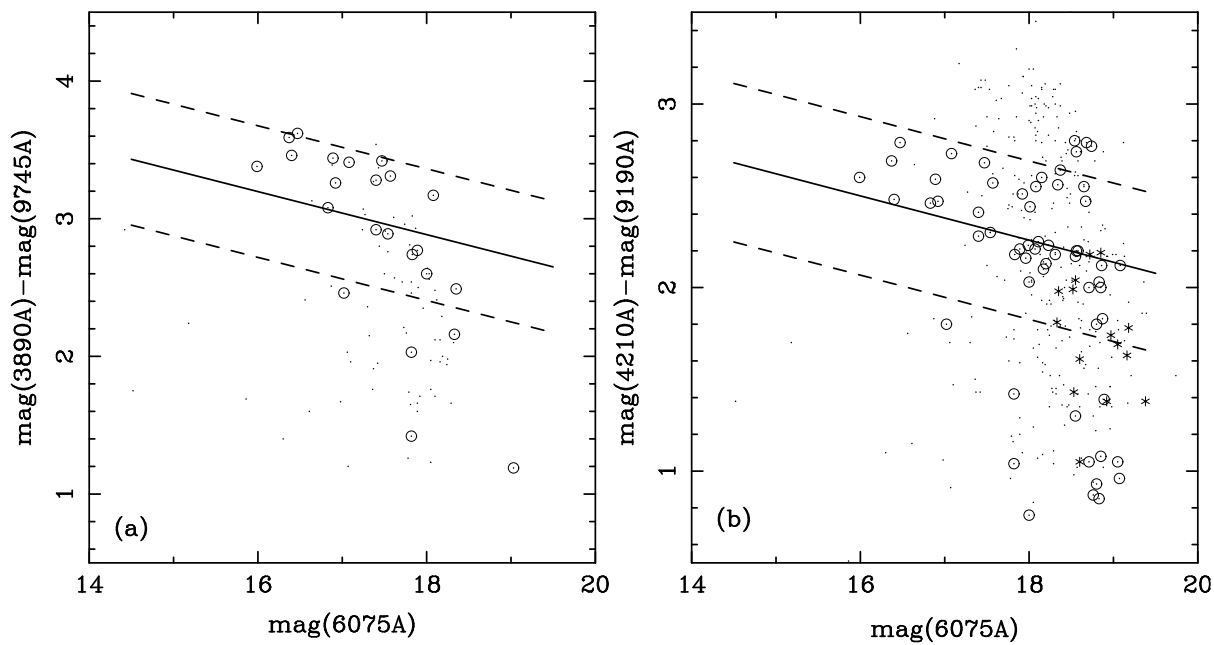}}
\caption{ The color-magnitude diagram for 359 faint galaxies. 68
early-type member candidates are denoted by ``$\odot$'', 8 spirals are
denoted by ``*'', and remainings by minor dots. The linear fits with
standard errors in color indices in Fig. 4 are also shown.}
   \label{Fig9)}  
\end{figure} 

\newpage 
\begin{figure}[h]
  \centerline{\psfig{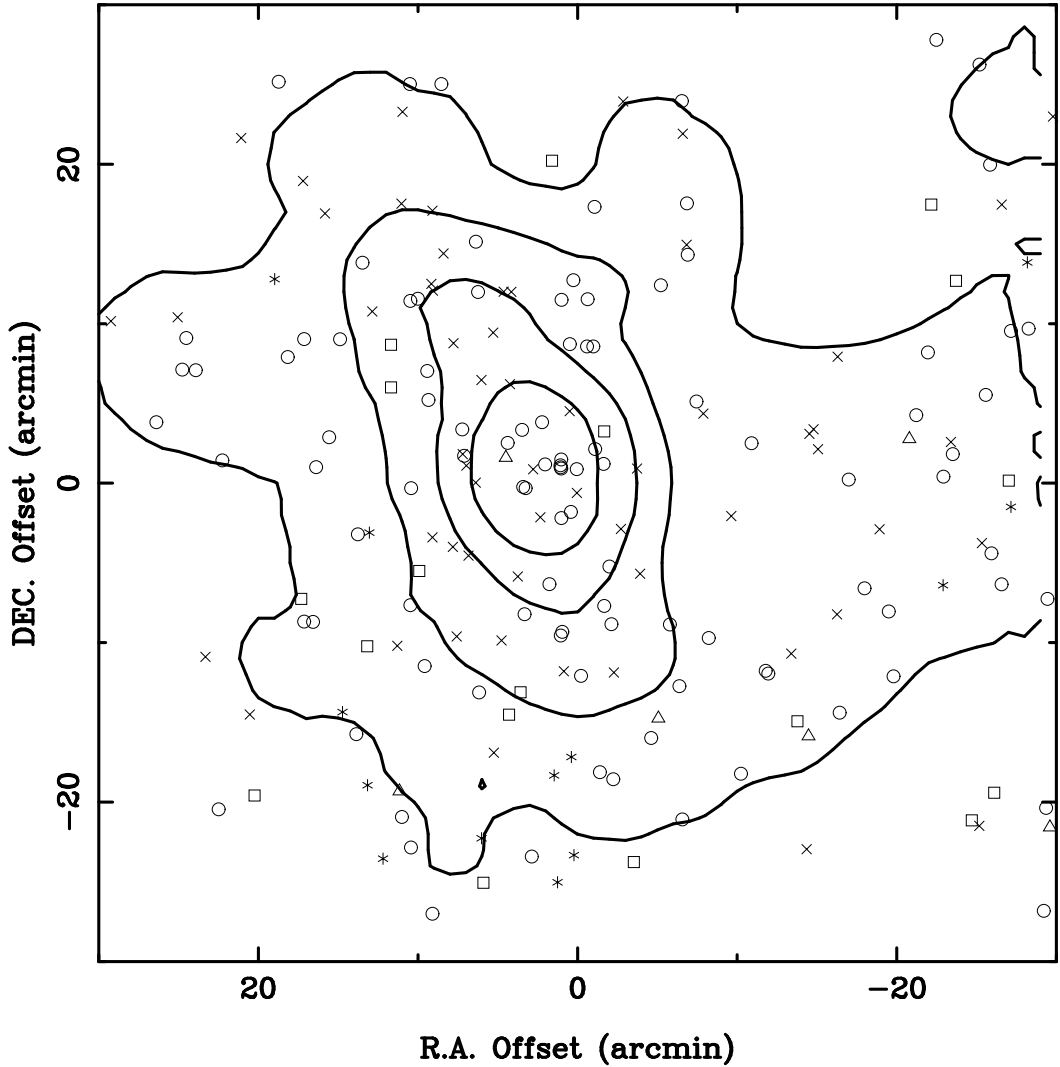}}
\caption{ The spatial distribution of all member galaxies, including 
102 ellipticals (``$\circ$''), 56 lenticulars (``$\times$''),13 Sa (``*''),
18 Sb (``$\Box$''), and 6 Sc (``$\triangle$'') galaxies. The contour map
of the surface density for all early-types using a $5 \times 5$ $arcmin^2$
smoothing window, is also given. The contour levels are 0.03, 0.08, 0.13,
0.18 $arcmin^{-2}$, respectively.}
\label{Fig10}  
\end{figure} 

\newpage 
\begin{figure}[h]
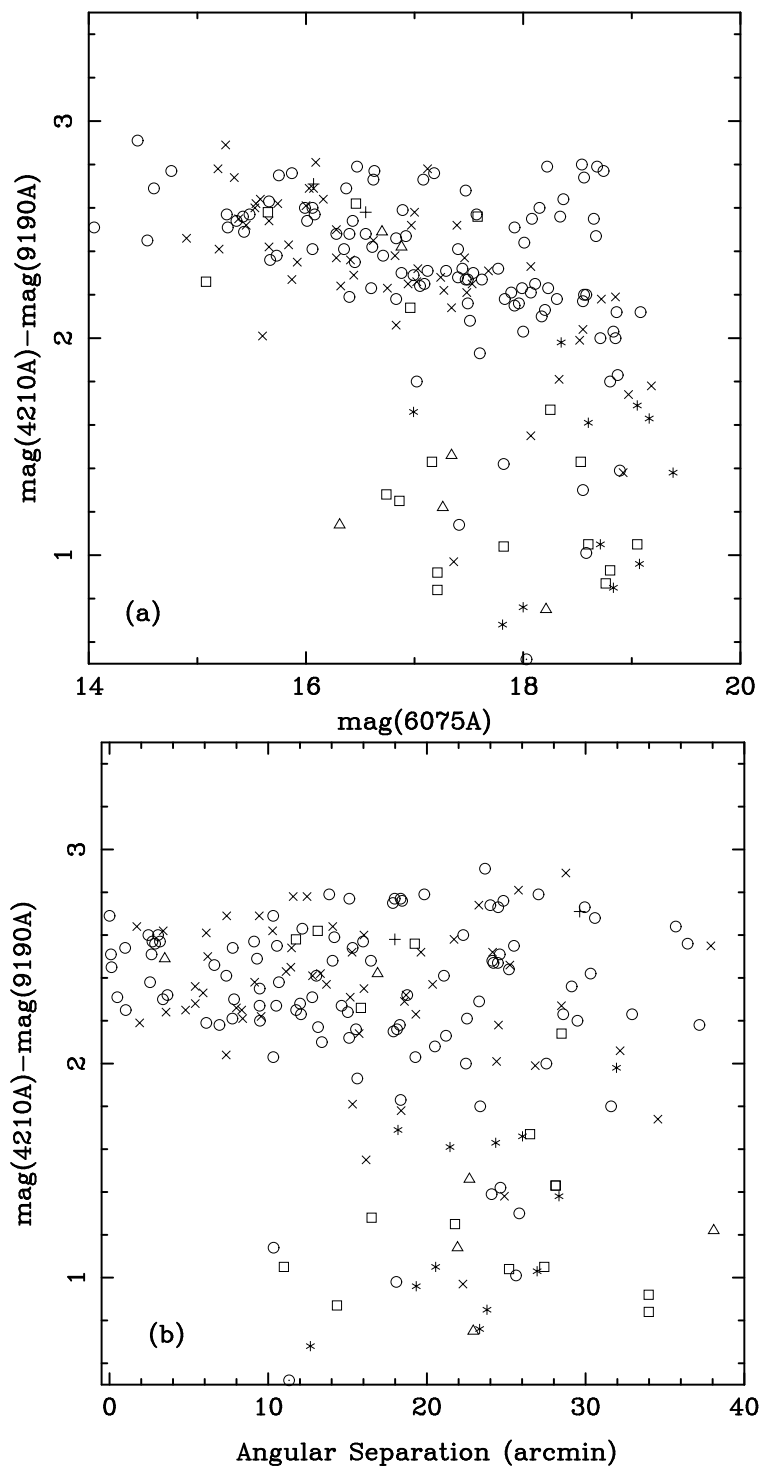

  \centerline{\psfig{figure=fig11a.ps,angle=270,width=9.9cm}}
  \centerline{\psfig{figure=fig11b.ps,angle=270,width=10.0cm}}
\caption{ (a) the color-magnitude relation for all member galaxies. 
(b) The plot of color versus angular  separation (in arcmin) for all member
galaxies. There are 198 member galaxies in our sample, including 102
ellipticals (``$\circ$''),  56 lenticulars (``$\times$''),2 S0a (``$+$''), 13
Sa (``*''), 18 Sb (``$\Box$''), 6 Sc (``$\triangle$''), and 1 Irr
(``$\odot$'')  galaxies. } 
\label{Fig11}  
\end{figure} 

\end{document}